\documentclass[14pt]{article}
\pdfoutput=1
\usepackage{amsmath,amssymb,amsfonts,amsthm,bm,bbm,cancel,wasysym}
\usepackage{epsfig,graphics,graphicx,epstopdf}
\usepackage{array,booktabs,colortbl,colordvi,multirow}
\usepackage{colordvi,color,xcolor}
\usepackage{hyperref}
\usepackage{rotating}

\usepackage{verbatim}
\usepackage{cite}
\usepackage{subfig}
\usepackage{setspace}
\usepackage{url}
\usepackage[percent]{overpic}
\usepackage{slashed}
\usepackage{authblk}
\usepackage{xspace}
\usepackage{fullpage}
\usepackage{hyperref}

\def\ie{{\it i.e.}}
\def\eg{{\it e.g.}}
\def\etc{{\it etc}}

\def\to{\rightarrow}

\newskip\zatskip \zatskip=0pt plus0pt minus0pt
\def\matth{\mathsurround=0pt}
\def\lsim{\mathrel{\mathpalette\atversim<}}
\def\gsim{\mathrel{\mathpalette\atversim>}}
\def\atversim#1#2{\lower0.7ex\vbox{\baselineskip\zatskip\lineskip\zatskip
  \lineskiplimit 0pt\ialign{$\matth#1\hfil##\hfil$\crcr#2\crcr\sim\crcr}}}

\begin{document}


\begin{flushright}
SLAC-PUB-251011\\
\today
\end{flushright}
\vspace*{5mm}

\renewcommand{\thefootnote}{\fnsymbol{footnote}}
\setcounter{footnote}{1}

\begin{center}

{\Large {\bf Identifying Charged Lepton-like Portal Matter at Future Colliders}}\\

\vspace*{0.75cm}

{\bf Thomas G. Rizzo}~\footnote{rizzo@slac.stanford.edu}

\vspace{0.5cm}

{SLAC National Accelerator Laboratory}\\ 
{2575 Sand Hill Rd., Menlo Park, CA, 94025 USA}

\end{center}
\vspace*{.5cm}


\begin{abstract}
\noindent  

In the Kinetic Mixing (KM) portal scenario, the interaction of dark matter (DM) with the particles of the Standard Model (SM) is generated by diagrams connecting the familiar photon with its dark 
sector analog, the dark photon (DP), via loops of particles carrying both dark and SM quantum numbers, \ie, Portal Matter (PM). For the case of sub-GeV DM and DP, these PM states may lie in the 
$\sim 1-10$ TeV range and be potentially accessible at the HL-LHC as well as at other future lepton and hadron colliders. In perhaps the simplest scenario of this kind, PM consists of just a pair of 
electrically charged, iso- and color-singlet, vector-like (VL) fermions having opposite dark charges, with an $O(1)$ mass splitting, yielding a finite value for the strength of the KM, \ie, 
$\epsilon \sim a~ few \times10^{-4}$. The dark Higgs induced mixing of PM states with their SM analogs allows for their decay but can also lead to significant distortions in the expected production 
properties for the PM at future lepton colliders due to $t$-channel dark Higgs exchange, potentially confusing PM identification. We show that the large set of clean observables available at lepton 
colliders is more that sufficient to resolve any of these ambiguities. The possibility of the production of like-sign PM fields via $t/u$-channel exchange of the same dark Higgs is also briefly explored.
\end{abstract}

\vspace{0.5cm}
\renewcommand{\thefootnote}{\arabic{footnote}}
\setcounter{footnote}{0}
\thispagestyle{empty}
\vfill
\newpage
\setcounter{page}{1}



\section{Introduction and PM Model Overview}

The Standard Model (SM) is extremely successful in explaining, or at least describing, a significantly large amount of experimental and observational data. However, we know that it cannot be the 
final story as it still leaves too many issues unaddressed. Although we know that dark matter (DM) exists and lies outside of the SM framework, its elemental nature remains 
completely mysterious.  For example, we don't know why the amount of it is not so different than that of the particles in the SM nor do we know whether or not in interacts with the SM in any 
non-gravitational manner.  If we assume that DM does not dominantly consist of primordial black holes\cite{Carr:2020xqk}, it would seem rather likely that for DM to achieve its observed relic 
abundance\cite{Planck:2018vyg}, an additional, yet to be observed, interaction of some kind between it and us is probably necessary. 

Historically, ideas related the identity of DM were correlated with solutions to problems with, or questions left unanswered by, the SM such as the solution of the strong CP problem, \ie, the 
QCD axion\cite{Kawasaki:2013ae,Graham:2015ouw,Irastorza:2018dyq}, and the hierarchy problem as addressed by,\eg, R-parity conserving supersymmety, \ie, the lightest neutralino as the 
archetypical  
example of a thermal Weakly Interacting Massive Particle (WIMP)\cite{Arcadi:2017kky,Roszkowski:2017nbc,Arcadi:2024ukq} generally thought to lie in the few GeV to $\sim 100$ TeV mass range. 
Such particles have been keenly sought since their conception decades ago but so far all of the searches have had negative results, ruling out substantial regions of the respective models' 
parameter spaces \cite{LHC,Aprile:2018dbl,Fermi-LAT:2016uux,Amole:2019fdf,LZ:2022ufs,PandaX:2024qfu,SuperCDMS:2024yiv,Aprile:2024xqi} although substantial regions still require 
exploration. These null results have led to an enormous growth in the set of possible DM scenarios with respect to its mass, as to how the observed relic abundance is obtained and how it can 
interact with the SM through the existence of new forces\cite{Alexander:2016aln,Battaglieri:2017aum,Bertone:2018krk,Cooley:2022ufh,Boveia:2022syt,Schuster:2021mlr,Cirelli:2024ssz}. This very 
large model space will clearly be experimentally challenging to explore and understand experimentally. Fortunately, effective field theories (EFTs) describing a large part of this model space at 
low energies have been developed, which can be either renormalizable or non-renormalizable and which describe with only a few parameters the specific form of the DM and its interactions with the 
SM, termed `portal models'. In addition to the DM itself, these models all require the existence of other new particles which act as mediators for these new interactions\cite{Lanfranchi:2020crw} but 
their exact nature depends upon the details of a particular model setup. 

As is well-known, the standard WIMP paradigm\cite{Steigman:2012nb,Steigman:2015hda,Saikawa:2020swg} which relies only upon SM interactions to mediate interactions with DM can only operate 
down to DM masses of a few GeV due to the Lee-Weinberg bound\cite{Lee:1977ua,Kolb:1985nn}. One of these attractive portal models allows us to extend this attractive idea to lower masses via the 
introduction of new forces.  In the simplest version of this renormalizable kinetic mixing (KM) - vector portal setupl\cite{KM,vectorportal,Gherghetta:2019coi}, a new $U(1)_D$ `dark' gauge interaction 
is introduced with a corresponding gauge boson, the dark photon (DP), $V$\cite{Fabbrichesi:2020wbt,Graham:2021ggy,Barducci:2021egn}. The DP couples to a conserved dark charge, $Q_D$, that 
is carried by the DM but not by the SM fields.  The $U(1)_D$ symmetry is generally broken, as in the discussion below, which then leads to a finite mass for the DP via the vevs of one or more 
dark Higgs (DH) fields\cite{Li:2024wqj} as in the SM. In some setups these same vevs can also contribute to the mass of the DM itself which here is assumed to lie in the same range as that 
of the DP, $\lsim 1$ GeV. At low energies, the interactions of dark sector fields with those of the SM are then generated by the kinetic mixing (KM) of the SM and DP at the 1-loop level via 
polarization-like diagrams. In order to generate such diagrams, however, other new fields, which are necessarily heavy, must also be present carrying both $U(1)_D$ dark 
as well as SM charges so that they can couple to both gauge bosons. Given their SM charges, for such new states to have avoided detection at accelerators up to now as well as to satisfy 
Higgs coupling, unitarity and precision electroweak constraints, these heavy fields must be either complex scalars and/or vector-like (VL) fermions 
 (VLFs)\cite{CarcamoHernandez:2023wzf,CMS:2024bni,Alves:2023ufm,Banerjee:2024zvg,Guedes:2021oqx,Adhikary:2024esf,Benbrik:2024fku,Albergaria:2024pji,Chen:2017hak,Biekotter:2016kgi}. 
We will refer to such exotic particles as Portal Matter (PM) and they have been the subject of much recent attention by ourselves and other authors\cite{Rizzo:2018vlb,Rueter:2019wdf,Kim:2019oyh,Rueter:2020qhf,Wojcik:2020wgm,Rizzo:2021lob,Rizzo:2022qan,Wojcik:2022rtk,Rizzo:2022jti,Rizzo:2022lpm,Wojcik:2022woa,Carvunis:2022yur,Verma:2022nyd,Rizzo:2023qbj,Wojcik:2023ggt,Rizzo:2023kvy,Rizzo:2023djp,Rizzo:2024bhn,Ardu:2024bxg,Rizzo:2024kzu,Rizzo:2025tap}. 

The detailed properties of these PM fields can, however, play an important role at and above the electroweak scale where such an EFT may have a desirable UV-completion. In particular, 
if the masses and couplings of these PM fields are known, these 1-loop diagrams can be explicitly calculated in-principle and the strength of this KM, described by a dimensionless parameter 
$\epsilon$, can be completely determined, \ie,  
\begin{equation}
\epsilon =\frac{g_D e}{12\pi^2} \sum_i ~(\eta_i N_{c_i}Q_{em_i}Q_{D_i})~ ln \frac{m^2_i}{\mu^2}\,.
\end{equation}
where $g_D$ is the $U(1)_D$ gauge coupling (and so we can also by analogy define $\alpha_D=g_D^2/4\pi$), $m_i(Q_{em_i},Q_{D_i}, N_{c_i})$ are the mass (electric charge, dark 
charge, number of colors) of the $i^{th}$ PM field, while $\eta_i=1(1/4)$ if the PM is a Dirac VLF (complex scalar).  Although $\epsilon$ is generally not finite, we might expect that it might be so in a 
UV-theory wherein group theoretical constraints will force the sum inside the parenthesis vanish, \ie, 
\begin{equation}
\sum_i ~(\eta_i N_{c_i} Q_{em_i}Q_{D_i})=0\,,
\end{equation}
making $\epsilon$ actually calculable in these modesl. In such a case we see that if $g_D \sim e$ and the PM fields were to be relatively close in mass we might then expect to find that 
$\epsilon \sim 10^{-(3-4)}$ which is roughly the same range as that needed for the DM to achieve its observed relic density and simultaneously satisfy other bounds from experiment and observation to 
which we turn.

There are numerous constraints that models of the present type, having a massive spin-1 mediator, must satisfy that arise from a host of various observations and experiments which include both 
direct and indirect detection in addition to accelerator searches as well as those from both astrophysics and cosmology. For example, to obtain the Planck abundance of DM the velocity-weighted 
annihilation cross section into SM particles at freeze-out must roughly satisfy the generic requirement\cite{Steigman:2012nb,Steigman:2015hda,Saikawa:2020swg}), 
$\sigma v_{rel}\sim 3\times 10^{-26}~$cm$^3$ sec$^{-1}$, with the exact value depending upon the details of the model. For DM in the mass range $\lsim 1$ GeV, during the time of the CMB this 
cross section must be suppressed by several orders of magnitude\cite{Planck:2018vyg,Slatyer:2015jla,Liu:2016cnk,Leane:2018kjk,Wang:2025tdx} to avoid adding additional 
electromagnetic energy into the thermal bath; quantitatively similar bounds must be also satisfied at present times\cite{Koechler:2023ual,DelaTorreLuque:2023cef,Wang:2025jhy}. This observation 
then implies that $\sigma v_{rel}$ must have a significant temperature ($T$) dependence which would {\it not} be the case if the DM in this setup were a Dirac VL fermion as the annihilation process 
would then be $s$-wave, hence, velocity and so temperature independent at leading order.  However, if the DM were instead a Majorana fermion or complex scalar, the corresponding annihilation 
cross section would instead be $p-$wave and so scale as $\sim v^2 \sim T$ which could more easily avoid such constraints{\footnote {See, however\cite{Belanger:2024bro}}}. However, even in this 
case, one finds that $p$-wave annihilation may be instead be constrained by considerations of the BBN\cite{Omar:2025jue}. A second possibility, if DM were to be pseudo-Dirac, is that the 
relic density is obtained via co-annihilation which can be a very strongly $T$-dependent process due to Boltzmann suppression if the mass splitting between the two dark states (due to, \eg, a $Q_D=2$ 
dark Higgs vev) is relatively large\cite{Brahma:2023psr,Balan:2024cmq,Garcia:2024uwf,Mohlabeng:2024itu}.  Simultaneously, if the mass splitting is sufficiently large, any constraints arising from 
either/both direct and indirect detection searches can also be rather easily avoided.

Interestingly, depending on the lower energy particle content of the specific model, \eg, in the case of pseudo-Dirac DM, it is possible to get a handle upon the high energy scale at which a 
UV-completion should take place - which is not far away. The reason for this is that, for a range of model parameters, the $U(1)_D$ gauge coupling, $\alpha_D$, via RGE evolution can grow to a 
non-perturbative value at a relatively low scale\cite{Davoudiasl:2015hxa,Reilly:2023frg,Rizzo:2022qan,Rizzo:2022lpm}. For example, in this setup, if the DM/DP masses are $\sim 100$ MeV and 
$\alpha_D\gsim 0.2$, the scale of the breakdown in perturbativity lies not far from $\sim 10-30$ TeV, before which we might expect that $U(1)_D$ should be embedded into some larger non-abelian 
gauge group, $G_D$, so that sign of the beta-function is flipped.  It is easy to imagine that the PM mass scale and that associated with the breaking of $G_D$ to $U(1)_D$ may be intimately related 
and this is commonly realized in PM model setups which have been the subject of much of our recent work\cite{Rizzo:2018vlb,Rueter:2019wdf,Rueter:2020qhf,Wojcik:2020wgm,Rizzo:2021lob,Rizzo:2022qan,Rizzo:2022jti,Rizzo:2022lpm,Rizzo:2023qbj,Rizzo:2023kvy,Rizzo:2023djp,
Rizzo:2024bhn,Rizzo:2024kzu,Rizzo:2025tap,Tewary:2025vij}.  A simple and now well-studied example of this is the `SM-like', $G_D=SU(2)_I\times U(1)_{Y_I}$\cite{Bauer:2022nwt} setup with the 
$U(1)_D$ embedded into $G_D$ exactly like $U(1)_{em}$ is in the case of the SM - except that it too is also broken but at the $\sim 1$ GeV mass scale{\footnote {This setup was originally motivated 
by earlier work on the $E_6$-type extended gauge models\cite{Hewett:1988xc}.}}. Of course, $G_D$ need not break down to $U(1)_D$ in a single step, but if so heavy gauge bosons 
analogous to the SM $W,Z$ (denoted as $Z_I,W_I$) would exist and obtain comparable large masses as part of this symmetry breaking process, leading to some interesting phenomenological 
possibilities at high energy colliders and elsewhere.

In this paper, we will explore the possibility that the PM takes the form of a pair of TeV-scale, VL charged fermions with an $O(1)$ mass splitting, which are both color and weak isospin singlets, 
qualitatively similar to the right-handed charged leptons, $\ell_R$, in the SM. The lighter of these two states can only decay, as is required by numerous constraints, if a small mixing occurs with the 
analogous SM state, the $\ell_R$, generated via the vev of a dark Higgs field with suitable quantum numbers. In the minimal scenario, this can be the same (and {\it only}) one as that which generates 
the DP mass thus breaking 
$U(1)_D$. As we will see below, if the SM charged leptons, $\ell$, with which the lightest PM fermion, $E$, admixes,  are the same as those that initiate the $\ell^+ \ell^- \to E^+E^-$ process at a 
future TeV-scale $e^+e^-$ or $\mu^+\mu^-$ collider, this dark Higgs interaction between $E$ and $\ell$ generates a new $t$-channel exchange diagram that can lead to significant modifications 
in the properties of the $E^+E^-$ pair production process that may mask the identity of this new heavy lepton-like particle as PM. Such new exchange processes will {\it not}, however, influence 
the production of these states at the HL-LHC or at any future hadron collider such as the FCC-hh at LO/NLO since there the initial states consists of only quarks and gluons. Here we will show, 
however, that lepton colliders allow for a sufficiently large set of precision observables to resolve any of these possible ambiguities and to establish the identity of the produced pair of charged, 
color-singlet fermions as PM with $Q_{em}=\pm 1$. 

The outline of this paper is as follows:  After this Introduction, in Section 2 we present the specific details of the PM model we will examine further wherein these PM fields are both color and isospin 
singlets but carry electric charge, $Q_{em}=-1$, and have essentially the same flavor as the colliding SM charged leptons in the collider initial state. To be as general as possible and, to cover the most 
challenging situation, we consider the case where these PM states have masses $\sim 1-2$ TeV, beyond current LHC bounds, but which also lie sufficiently far below the scale of the full UV-completion 
with its associated additional gauge fields, \etc, that the effects of these more massive states can be ignored for the practical purposes considered here. 
In Section 3, we discuss the various observables that can be employed at lepton colliders to distinguish the production of PM states from that of the more commonly discussed 'ordinary' VL 
charged leptons\cite{Shang:2021mgn,Guo:2023jkz,Yue:2024sds} including the angular and energy dependence for their production and longitudinal and/or transverse polarization asymmetry 
observables. We demonstrate that this set of observables is sufficiently rich, given the anticipated level of statistics and ease at which the signal can be distinguished from SM backgrounds, that such 
a separation can be easily made.  Of course, these tentative conclusions should be verified by future detailed simulations. 
In Section 4 we briefly consider the production of like-sign VL PM fields via $t/u-$channel of the same dark Higgs exchange.  This process is a necessary result of the 
lepton-like, VL PM setup that we consider here and is relatively SM background-free.  As will be seen, while this cross section can be large, its magnitude depends quite sensitively on the value of the 
dark Higgs Yukawa coupling. Finally, a discussion and our conclusions can be found in Section 5.


\section{Basic Model Generalities}

Perhaps the simplest, renormalizable dark sector we can imagine given the above discussion, consists of just the familiar $U(1)_D$ gauge field, 
the DP ($V$), whose mass is generated by the $\lsim$ GeV scale vev of a single dark Higgs, $h_D$, carrying $Q_D \neq 0$. We also require that there be a pair of, in the present case, 
VLF PM fields with the same SM quantum numbers (\eg, color and electric charge) and 
similar TeV-scale masses but with opposite values of $Q_D$ to render the KM parameter, $\epsilon$, introduced above, finite and in the phenomenologically desired range.  Of course, 
these PM fields must necessarily be unstable to avoid numerous experimental, astrophysical and cosmological constraints and so the dark Higgs field must play an important secondary role. 
As has been previously discussed\cite{Rizzo:2018vlb,Rizzo:2022qan}, to allow for the decay of lightest PM fermion carrying a non-zero value of $Q_D$, a mixing must occur with an analogous 
SM field carrying 
the same color and electric charge via a dark Higgs field with the corresponding quantum numbers to maintain gauge invariance. This requirement usually limits us to consider both the VLF PM as 
well as the dark Higgs field to be either in SM weak isosinglets or isodoublets. For example, in the case of colorless, lepton-like PM VL fermions which will concern us below, we can imagine their SM 
representations taking the generic forms (here the primed and unprimed $E$ fields are distinct) 
\begin{equation}
E_{L,R},~~ \begin{pmatrix} N \\ E'\\ \end{pmatrix}_{L,R}\,,
\end{equation}
so that one can further imagine that the possible PM-SM mixing terms with the following structures could be generated: 
\begin{equation}
 {\cal L}_{mix} = ~ y_S \bar E_L\ell_Rh_{D_S} +y_S'(\bar \nu,\bar \ell)_L^T(N,E')_R^T h_{D_S}^\dagger+y_D(\bar \nu,\bar \ell)_L^TE_Rh_{D_D}^\dagger+y_D' (\bar N,\bar E')_L^T \ell_R h_{D_D}+h.c. \,
\end{equation}
where $h_{D_{(D,S)}}$ here represents a weak $SU(2)_L$ isodoublet or isosinglet dark Higgs field, respectively, and the $y$'s are assumed to be distinct $O(1)$ Yukawa couplings. Note that these 
interactions are individually all chiral. We further see that if both isosinglet and isodoublet dark Higgs were to be  {\it simultaneously} present with comparable vevs and Yukawa couplings then 
either (or both) of $E,E'$ will be able to mix with the SM charged lepton with both chiral structures. Here we have 
dropped all reference to possible lepton flavor labels so that $N,E^{(')}(\nu,\ell)$ will just represent PM (SM) generic leptons; all flavor issues will be ignored for simplicity in the discussion below. 
We again note that the primed and unprimed labels in this expression are being used to distinguish the PM fields with differing $SU(2)_L$ quantum numbers{\footnote {In a similar fashion, it is 
easily imagined that analogous interaction terms can also be straightforwardly written down for other, \eg, quark-like PM quantum number choices.}}.  Of these possibilities, certainly the least 
complex choice corresponds to just the first term in the expression above where both $E$ and $h_D$ are weak isosinglets and this is the setup that we will generally assume in the discussion that 
follows, \ie, the isosinglet PM scenario. A more UV complete model of this kind has recently been constructed\cite{Rizzo:2025tap} wherein this simple setup is obtained in the low energy limit. 

Once the dark isosinglet Higgs obtains a vev, $v_s$,  we can decompose this dark Higgs as $h_{D_S}= (v_s +S +iG_V)/\sqrt 2$, where $G_V$ is just the Goldstone boson eaten by $V$ while the 
real field, $S$, remains in the physical spectrum (and which is also frequently termed the dark Higgs). As is easily seen, this interaction term in ${\cal L}_{mix}$ will now allow for the PM decays of 
the form $E\to e(S,V=V_{long}\simeq G_V)$, with rates controlled by the $O(1)$ values of the Yukawa coupling, $y_S=y$, in the discussion below. Clearly, the Goldstone Boson Approximation 
(GBA)\cite{GBET} limit is applicable here as $m_{S,V}^2 << M_E^2$ since in all cases the PM fermion masses must be $\gsim 1$ TeV or more in order to have so far evaded experimental 
detection at the LHC. 
It is easy to see that in this same Goldstone limit $\Gamma(E\to eS)=\Gamma(E\to eV)$ since $m_e^2<<m_E^2$  and $V\simeq G_V$, so that in this same approximation one explicitly finds that 
\begin{equation}
\Gamma(E\to eV,~eS)\simeq \frac{y^2 M_E}{32\pi}\,,
\end{equation}
which, as was previously discussed\cite{Rizzo:2018vlb,Rizzo:2022qan}, will be the by-far dominant decay mode(s) for $O(1)$ values of $y$; note that this result will remain true even for values as 
small as $y\lsim 0.1$ or so. 

\begin{figure}[htbp]
\centerline{\includegraphics[width=5.0in,angle=0]{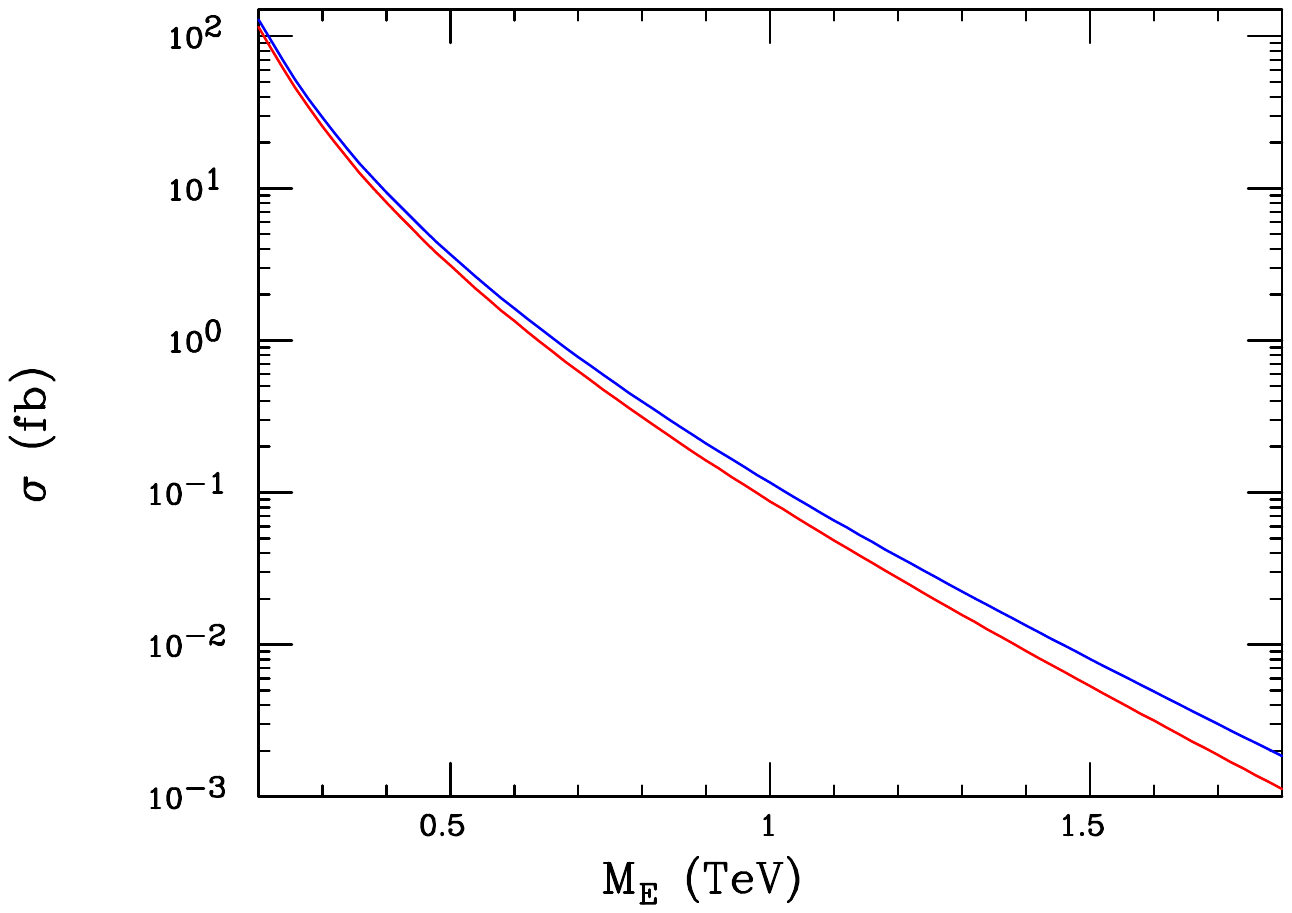}}
\vspace*{-0.8cm}
\centerline{\includegraphics[width=5.0in,angle=0]{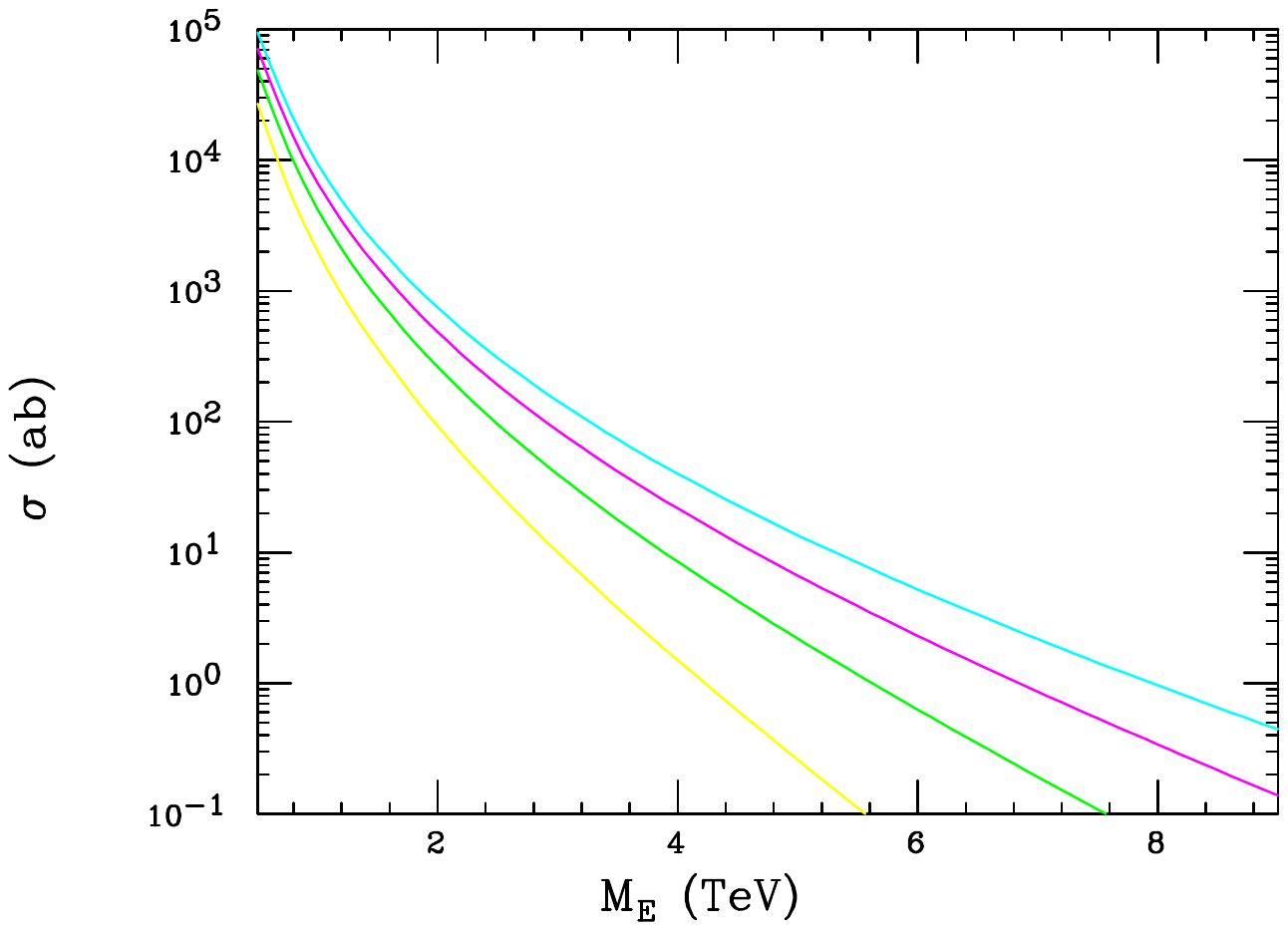}}
\vspace*{-1.3cm}
\caption{Cross sections for the pair production of SM isosinglet vector-like charged leptons, $E$, at present and future hadron colliders as functions of their mass: (Top) at the $\sqrt s =13(14)$ TeV 
LHC corresponding to the bottom red(top blue) curve; (Bottom) Similarly at the FCC-hh assuming that, from top to bottom, $\sqrt s=100(80,60,40)$ TeV, respectively. }
\label{fig1}
\end{figure}

The current lower bound on the mass, $M_E$, when the isosinglet state $E$ decays to electrons, under the assumption that $V,~S$ will appear in the detector as missing transverse energy (MET), 
(in a manner similar to charged slepton production\cite{Freitas:2003yp} followed by decay to the LSP) has been obtained by the authors of Ref.\cite{Guedes:2021oqx}, $\simeq 0.90$ TeV,  by 
the recasting of 139 fb$^{-1}$ of 13 TeV data from the LHC, employing as input several analyses from both 
ATLAS and CMS searches. We note that in the case where $E$ decays instead to muons the corresponding bound from the same set of analyses is only slightly weaker, \ie, $\simeq 0.85$ TeV. 
For reference, the cross section for the production of VL isosinglet $E^+E^-$ pairs at the $\sqrt s=13(14)$ TeV LHC is shown in the upper panel of Fig.~\ref{fig1}. 
These same authors have also extrapolated these results, in the case of decay to electrons, to obtain possible future exclusion (not discovery!) reaches at the HL-LHC with an luminosity of 
3 ab$^{-1}$ of 1.45 TeV (but still assuming that $\sqrt s=13$ TeV) and at the $\sqrt s=100$ TeV FCC-hh, also with an assumed integrated luminosity of 3 ab$^{-1}$, of 3.33 TeV.  An increase in 
the FCC-hh integrated luminosity by an order of magnitude may be able to push this mass limit of the FCC-hh up into the neighborhood of $\simeq 4.5$ TeV whereas a reduction of $\sqrt s$ 
to 40(60,80) TeV may result in a corresponding decreased reach of roughly $\simeq 55(30,13)\%$ or so given the falling cross section as indicated by the lower panel in Fig.~\ref{fig1}.

It is important to recognize that these results have been obtained under the assumption, as is usually made, that the relevant cross section in LO/NLO arises only from the familiar SM $\gamma,Z$ 
$s$-channel exchanges for the (sub)process $q\bar q \to E^+E^-$. Within the SM, it is easy to convince oneself that gauge boson fusion contributions to the production of these states is relatively 
suppressed as they are VL (thus having no axial-vector to enhance couplings to the SM $Z$) and are also weak isosinglets (thus not having any couplings to the SM $W$).  However, depending 
upon how 
this simple low energy model is UV completed, other exchanges may frequently be present - in particular, the $s$-channel exchange of additional new neutral gauge bosons. Typically, and certainly 
in the case of the well-studied $G_D=SU(2)_I \times U(1)_{Y_I}$ scenario\cite{Rueter:2019wdf}, the $E$ fermion has both vector as well as axial-vector couplings to the additional neutral gauge field, 
$Z'=Z_I$, generally 
distorting the predictions of the SM if the $Z'$ is relatively light. In the discussion below, since these aspects of the present scenario are highly model-dependent, we will concentrate on just the 
minimal model particle content described above and as realized in the low energy limit of the model in Ref.\cite{Rizzo:2025tap}; for practical purposes this corresponds to sending the masses of 
any of these potential new neutral gauge bosons to infinity relative to the $\sim$ TeV energy scales that we will be considering below so that their contributions to the $E^+E^-$ production process will 
decouple. We should, however, be mindful that such new contributing, $Z'$-like fields and more general $s$-channel exchanges may alter the simple setup that we will consider here; however,  if 
so, such fields will likely then be sufficiently light so as to announce their existence on their own such ass appearing as resonances in other final state channels such as Drell-Yan production. In 
any case, as we will note, their signatures will appear to be quite different than those of the scenario we consider below.

Within the rather simple effective framework above, consisting of the SM, the DP and the associated dark Higgs, plus the VL isosinglet, lepton-like PM field{\footnote {For further simplicity in the 
discussion below, we will assume that the pair production of the {\it heavier} VL PM field is beyond the kinematic reach of the relevant lepton collider. In the benchmarks that we consider below this 
would correspond to the second PM VL lepton being only roughly $\sim 40-50\%$ or so more massive.}}, there still remains some additional 
{\it light} exchanges in the production of the $E^+E^-$ final state that we have so far ignored and which, in the case of this lepton-like PM, will {\it not} influence any of the searches for such particles 
at hadron colliders. These involve the necessary $\bar E_L\ell_Rh_D+h.c.$ couplings discussed above that allows for the PM field $E$ to decay. For example, if $E$ is electron-like, the 
process $e^+e^- \to 
E^+E^-$ not {\it only} involves the SM $s$-channel $\gamma,Z$ exchanges but also the $t$-channel exchange of the light dark Higgs field which, as we'll see below, can be very important if the Yukawa 
coupling, $y$, is at all sizable. If $E$ were instead muon-like, the corresponding $t$-channel contribution would then be relevant at TeV-scale muon colliders. Obviously, however, since our PM is 
lepton-like, such exchanges will be absent from the corresponding $q\bar q$ initiated production process relevant at hadron colliders.

In what follows we will consider the generic future lepton collider production process, $\ell^+ \ell^- \to E^+E^-$, with $\ell=e,\mu$, even though we will continue to make use of the generic PM $E$, 
\etc, labels employed above.

\section{Observables at Lepton Colliders}

When considering the nature of the production cross section of any charged VL fermion at $\ell^+\ell^-$ colliders assuming only SM exchanges, one notes the absence of any axial vector 
coupling of $E$ to the SM Z. This leads to characteristic features of the cross section, like the absence of a Forward-Backward Asymmetry, $A_{FB}$. Indeed, 
one might use the fact that, since $A_{FB}=0$ is an expected hallmark of this production in such cases, an observed {\it non-zero} value for $A_{FB}$ would be potentially indicative of the production 
of particles which {\it do} have such axial vector couplings and so are {\it not} VL. To this end it important to examine how the presence of the $y\bar E_Le_Rh_{D_S}$-type coupling, now contributing 
as part of the $E^+E^-$ production process, would potentially alter these simple expectations in the case of PM, how this may lead to a possible confusion in identifying the true nature of the PM field 
$E$ and how this confusion can be resolved - a subject to which we now turn. For comparisons, we sometimes have in mind not just a generic isosinglet VL but also a possible isodoublet, as well as a 
fourth generation heavy lepton. Further, we'd also like to differentiate our PM isosinglet from the case where a heavy $Z'$ may be present and where the heavy lepton candidate may be VL with respect 
to the SM but is not so with respect to the extended gauge group leading to this $Z'$.

\subsection{Cross section and Forward-Backward Asymmetry}

Since, as mentioned above, current LHC direct searches place a lower bound on the mass of the PM lepton $E$, $M_E \gsim 0.9$ TeV, $E^+E^-$ production will require a future lepton collider 
having $\sqrt s \gsim 2$ TeV or so and thus will probe the region far beyond the reach of FCC-ee. 
Thus, since in such a case $s>>m_V^2, m_S^2 \lsim 1$ GeV$^2$, we should expect that the GBA to hold to quite high accuracy so that we can treat the $t$-channel 
exchange arising from the $y\bar E_Le_Rh_{D_S}$ coupling as just that due to an approximately massless complex dark Higgs scalar. 
Under such an assumption, for unpolarized and massless leptons in the initial state, the differential cross section for the $\ell^+ \ell^- \to E^+E^-$ process is given by (with the repeated indices here 
summed over as usual) 
\begin{equation}
\frac{d\sigma}{dz}=\frac{\pi\alpha^2}{2s}\beta~\Bigg(P_{ij}B_{ij}[2-\beta^2(1-z^2)] +2\lambda R_iC_i ~\Big[\frac{(1-\beta z)^2+1-\beta^2}{a-\beta z}\Big]+ \Big[\frac{\lambda}{2}~ \frac{(1-\beta z)}{(a-\beta z)}\Big]^2 \Bigg)\,,
\end{equation}
where $z=\cos \theta$, the production angle of $E^-$ with respect to the incoming negatively charged lepton. Here, with the mass of the $t$-channel exchanged dark Higgs has  been neglected, 
$\beta^2=1-4M_E^2/s$ and $a=(1+\beta^2)/2$. We have also defined 
\begin{equation}
P_{ij}=s^2 ~\frac{(s-m_i^2)(s-m_j^2)+(m_i\Gamma_i)(m_j\Gamma_j)}{[(s-m_i^2)^2+(m_i\Gamma_i)^2][i \to j]},~~~R_i=\frac{s(s-m_i^2)}{(s-m_i^2)^2+(m_i\Gamma_i)^2}\,,
\end{equation}
which are the familiar propagator functions together with the coupling combinations 
\begin{equation}
B_{ij}=(v^e_iv^e_j+a^e_ia^e_j)v^E_iv^E_j,~~~C_i=(v^e_i-a^e_i)v^E_i\,,
\end{equation}
where $m_i,\Gamma_i, (i=1,2)$ are the masses and decay widths of the $i^{th}$ gauge boson, \ie, the photon and $Z$, respectively, $v^{e,E}_i,a^e_i$ are correspondingly the vector and axial vector 
couplings of the initial lepton and VL PM fermions in units of the positron's charge, $e$, to these gauge bosons, \ie, $v^{e,E}_1=-1$, $a^e_1=0$, $a^e_2=-1/(4s_wc_w)$, \etc, with 
$s_w(c_w)=\sin \theta_w(\cos \theta_w)$ as usual, and where we have explicitly taken $a^E_i=0$. Furthermore, we have also defined the coupling ratio $\lambda =(y/e)^2$, which can be a 
relatively sizable parameter since $e\simeq 0.31$; as we shall see below, even a relative small value of $y$ can lead to significant distortions in the naive expectations for the pair production 
of isosinglet VL leptons.

The most obvious thing that we notice from the expression above is that, as well-known, the purely $s$-channel terms are even functions of $z$ since $E$ is indeed VL within the SM context. 
However, both the $s-t$-channel interference term as well as the pure $t$-channel exchange term lead to an angular asymmetry in the production cross section due to the forward, dark Higgs 
induced $t$-channel pole. Clearly, if the magnitude of $\lambda$, \ie, $y$, becomes significant, these deviations away from our naive `VL lepton' expectations that are based on the SM 
production channels alone will grow accordingly. 

\begin{figure}[htbp]
\centerline{\includegraphics[width=5.0in,angle=0]{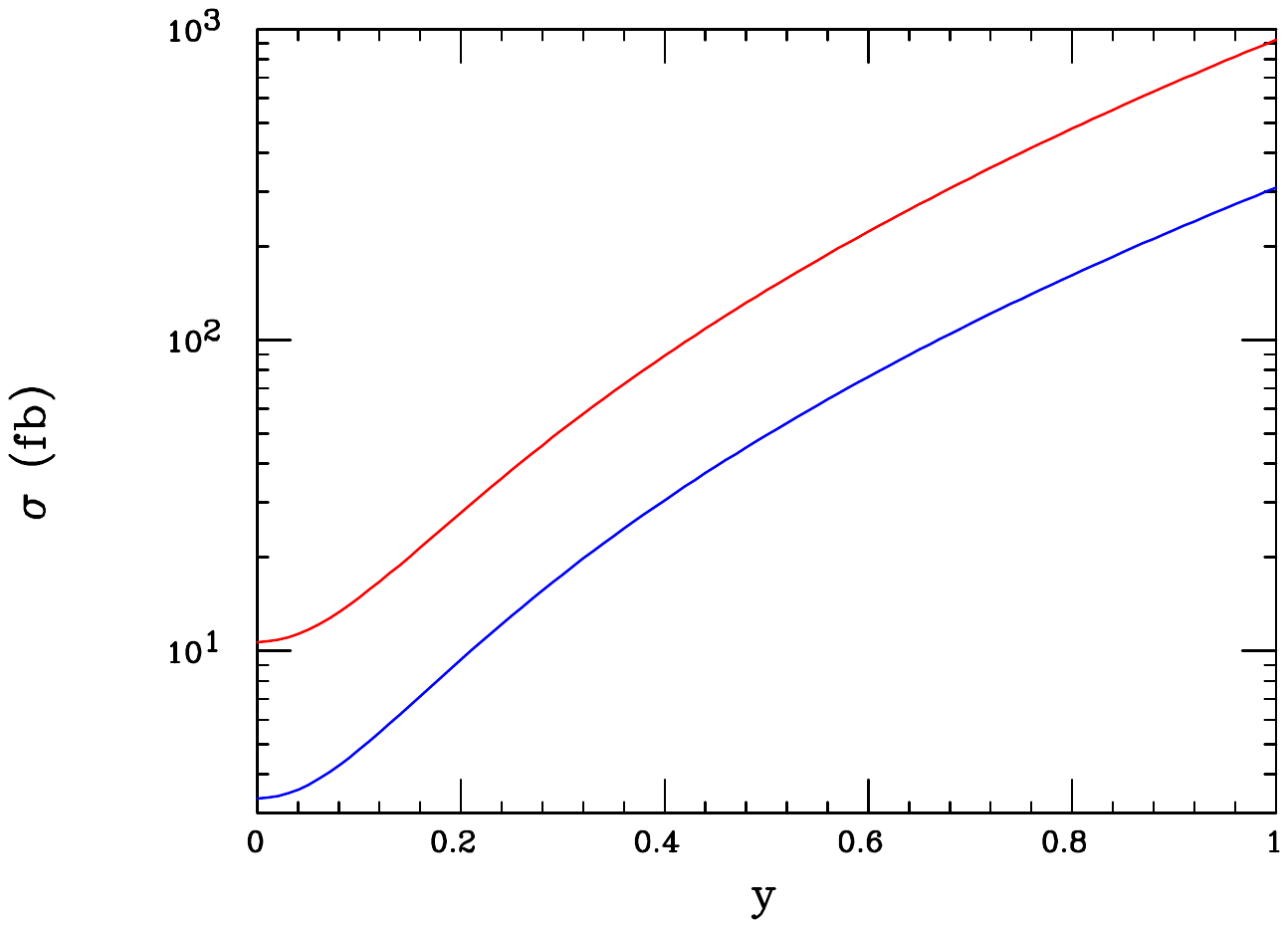}}
\vspace*{-0.8cm}
\centerline{\includegraphics[width=5.0in,angle=0]{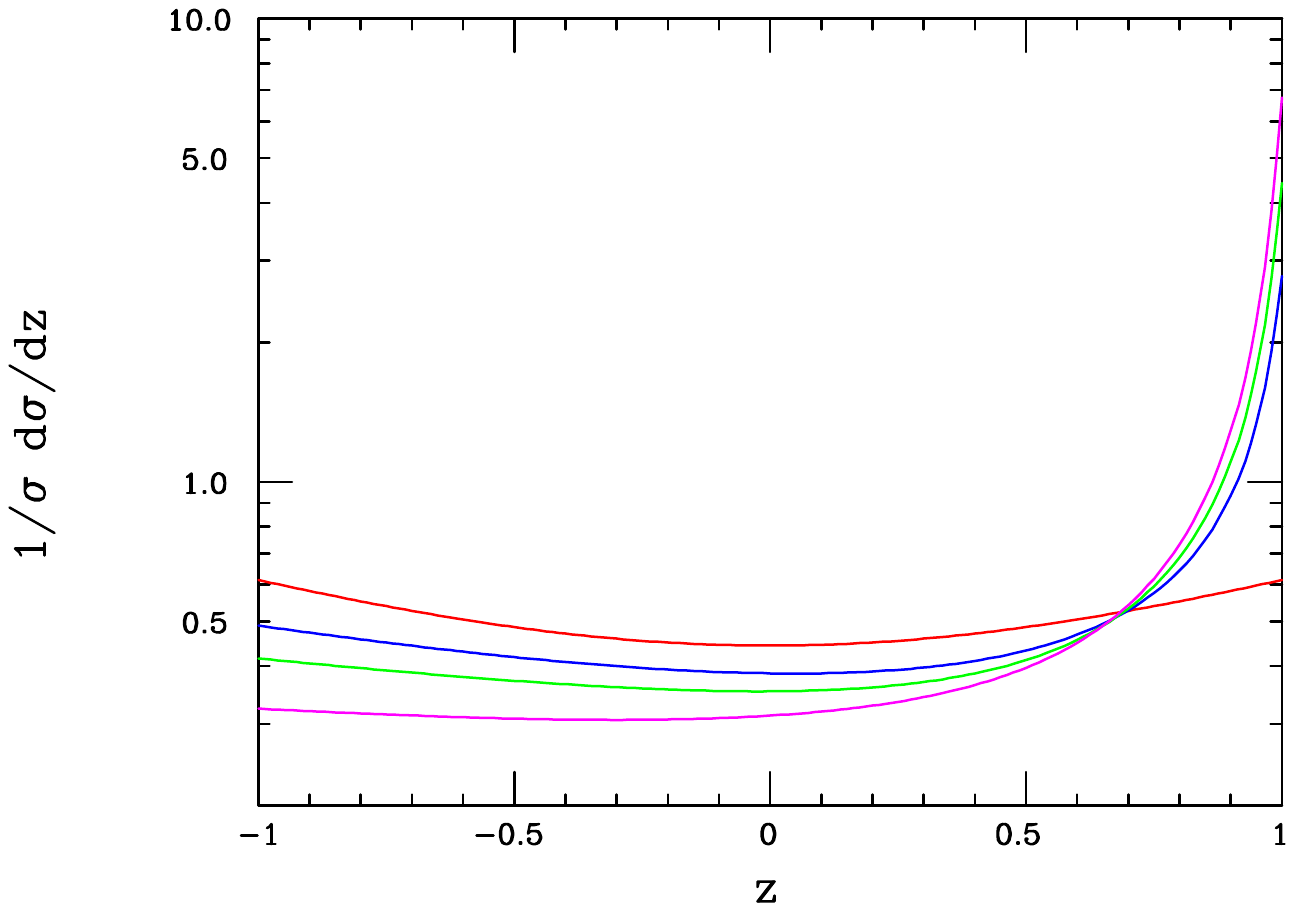}}
\vspace*{-1.3cm}
\caption{(Top) Total production cross section, $\sigma$, in $\ell^+\ell^-$ annihilation for pairs of vector-like, isosinglet PM, $E$, as a function of the Yukawa coupling parameter $y$, 
defined in the text, assuming that either BM1 with 
$\sqrt s=3$ TeV with $M_E=1$ TeV (top red) or BM2 with $\sqrt s=5$ TeV with $M_E=2$ TeV (bottom blue), respectively. (Bottom) The normalized angular distribution, as a function of $z$ as defined in 
the text, for PM production in $\ell^+\ell^-$ annihilation at $\sqrt s=3$ TeV with $M_E=1$ TeV assuming that  $y=0, 1/4,1/2,1$, corresponding to the curves from top to bottom on the left-hand side 
of the plot in red, blue, green and magenta, respectively.}
\label{fig2}
\end{figure}

As a more specific example of the dark Higgs exchange induced effects, we see in the upper panel of Fig.~\ref{fig2}, the $z$-integrated, total $E^+E^-$ production cross section, $\sigma$, as a 
function of $y$ for two representative benchmark points for $m_E=1,2$ TeV with$\sqrt s=3,5$ TeV, respectively. To be specific, for this and later use we define the benchmarks
\begin{enumerate}
\item[] \textbf{BM1}: $m_{E}=$ 1 TeV, $\sqrt{s}=$ 3 TeV
\item[] \textbf{BM2}: $m_{E}=$ 2 TeV, $\sqrt{s}=$ 5 TeV
\end{enumerate}
In this Figure, one observes that even for rather small values of $y\lsim 0.1-0.2$ there 
is a quite significant cross section enhancement 
which we see only becomes even more important as the value of $y$ increases further due to the presence of the approximate $t$-channel pole, something not expected in a generic model with 
VL leptons.  Such a behavior would differentiate this source of a non-zero $A_{FB}$ from that due to, \eg, an additional new heavy $Z'$ exchange in the $s$-channel under whose corresponding 
gauge symmetry the heavy lepton is {\it not} VL as occurs in some more UV-complete dark sector models such as the $G_D$ above The lower panel in this same Figure shows the 
normalized angular distribution, $\sigma^{-1}d\sigma/dz$ as a function of $z$, for the lighter benchmark point BM1 for various assumed fixed values of $y$. When $y=0$ we see the familiar 
$z\to -z$ symmetric angular distribution as naively expected for the production of a VL lepton; however, as $y$ 
increases more and more of the cross section is pushed into the forward hemisphere due to the dark Higgs exchange induced $t$-channel pole. Obviously, we see from this that we might anticipate 
that as $y$ increases from away zero, the deviations from the naive expectations for VL, isosinglet lepton production for other observables will also become quite significant. This is indeed what we 
will find below. 

Note that with the benchmark cross sections in the $\sim 3-10$ fb range and above, even relatively modest integrated luminosities, \eg, ${\cal L}=$ 5 ab$^{-1}$, will yield large samples of events,  
$\gsim 10^4$, so that statistics is not generally an issue in performing studies of these final states. SM backgrounds, however, mostly arising from $W^+W^-$ pair-production followed by same 
flavor leptonic decay are important, but will globally appear quite different since the $W$'s are significantly boosted compared to the PM fields, \ie, for the first benchmark model (BM1) where 
$M_E=1$ TeV and $\sqrt s=3$ TeV, $\gamma_W \sim 20$ while $\gamma_E \sim 1.5$; the difference is even larger for the case of the second benchmark model. This implies that in the 
$W^+W^-$ case, the same-flavor charged leptons in the final state are quite close to being back-to-back, essentially in the $W^\pm$ direction, and that the amount of missing transverse energy 
(MET) is relatively small as the two neutrinos will approximately balance each other. The PM final state, on the other hand due to the lower boost, will instead yield relatively uncorrelated, 
non-back-to-back leptons with a significant MET since the 
angular distribution for the decay is flat in the PM rest frame. Other backgrounds coming from, \eg, $ZZ$ production, can be significantly reduced by simply removing events where the observed 
dilepton invariant mass is close to that of the $Z$ while a jet veto will be very effective at removing other possible SM final states involving quarks or complex cascades.

Of course as is well-known, in the case that longitudinal lepton polarization is available, much of this $W^+W^-$ background, particularly in the forward direction, can be further significantly reduced (by 
up to a factor of $\sim 30$ or more) using a simple judicious choice of the, \eg, $e^\pm$ polarizations without much influence on the PM pair-production signal cross section.

\begin{figure}[htbp]
\centerline{\includegraphics[width=5.0in,angle=0]{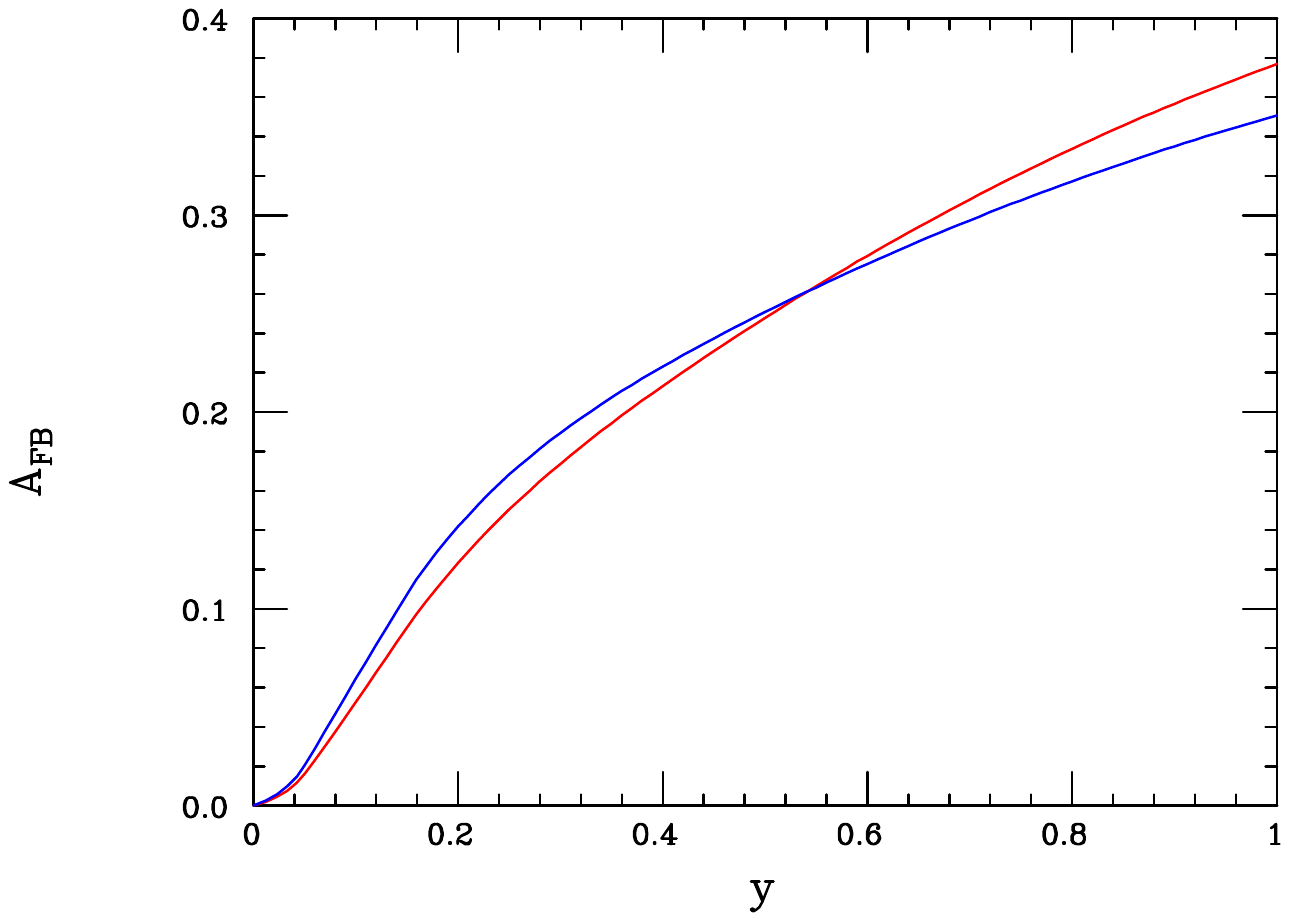}}
\vspace*{-0.8cm}
\centerline{\includegraphics[width=5.0in,angle=0]{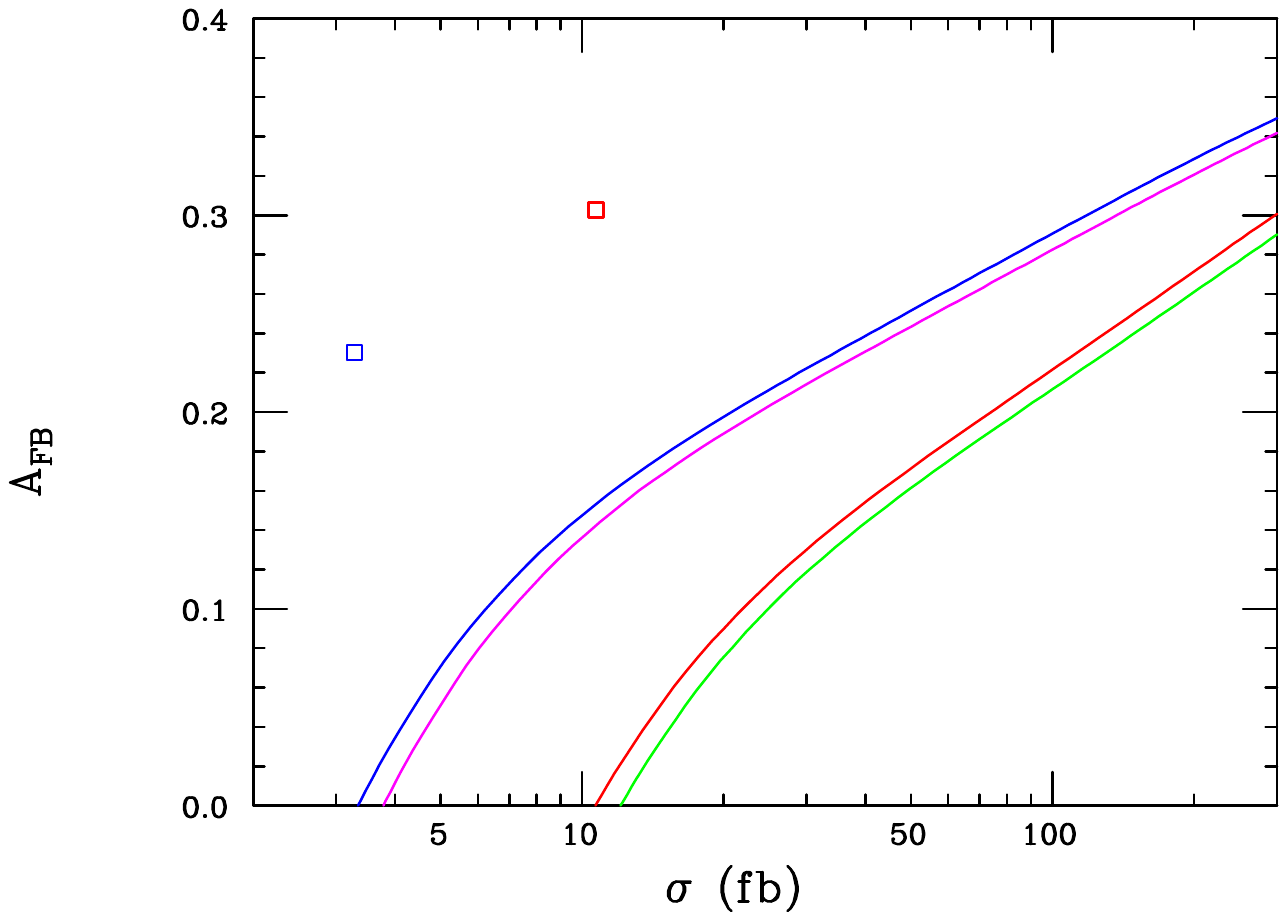}}
\vspace*{-1.3cm}
\caption{(Top) Angular averaged value of the forward-backward asymmetry, $A_{FB}$, for the $\ell^+ \ell^-$ production of pairs of isosinglet VL PM as a function of the Yukawa coupling $y$, as defined 
in the text, assuming that either 
$\sqrt s=3$ TeV with $M_E=1$ TeV (BM1 bottom red) or $\sqrt s=5$ TeV with $M_E=2$ TeV (BM2 top blue) curve, on the left-hand side of the plot, respectively. (Bottom) Correlation between the total 
production cross section, $\sigma$ and the angular averaged forward-backward asymmetry, $A_{FB}$, with the value of $y$ varying along the curves, assuming $\sqrt s=3$ TeV with $M_E=1$ 
TeV (BM1 top) or $\sqrt s=5$ TeV with $M_E=2$ TeV (BM2 bottom) pair of curves. The upper (lower) curve in each pair is for isosinglet (isodoublet) VL PM for purposes of comparison. The 
small upper (red) 
and lower (blue) squares are the corresponding predicted values of these same observables for a chiral, fourth general charged lepton, $L$, assuming  $\sqrt s=3$ TeV with $M_L=1$ 
TeV (BM1-like upper) or $\sqrt s=5$ TeV with $M_L=2$ TeV (BM2-like lower), also shown for comparison purposes.}
\label{fig3}
\end{figure}

Based on these considerations and to quantify this non-trivial angular dependence, we can as usual define the $z$-independent quantity, the forward-backward asymmetry, $A_{FB}$, by first 
forming the ratio of differential cross sections
\begin{equation}
R=\frac{d\sigma(z\geq 0)-d\sigma(z\leq 0)}{d\sigma(z\geq 0)+d\sigma(z\leq 0)}\,,
\end{equation}
and then integrating both the numerator and denominator of this expression over the relevant ranges of $z$. Needless to say, we expect $A_{FB}$ to be strongly $y$-dependent and to vanish when 
$y=0$ as it would for a canonical VL lepton; these expectations are borne out by viewing the top panel in Fig.\ref{fig3} where we see that $A_{FB}$ grows rapidly away from zero as the value of $y$ 
increases due to the forward pole produced by the dark Higgs $t$-channel exchange.

Taking both the total cross section, $\sigma$, together with $A_{FB}$ we can obtain a correlation plot where the value of $y$ grows from left to right along the respective curves; this is seen 
in the lower panel of Fig.~\ref{fig3}. Combining these measurements should provide sensitivity down to very small values of $y\lsim 0.02-0.03$ or so with the above assumed integrated luminosity.  
Here we see not only the results for the two benchmark points but {\it also} the analogous ones we'd obtain if the VL PM was instead the lower member of an isodoublet as described above. These two 
sets of results are qualitatively similar but would be easily separated by the high precision data expected at a $\ell^+ \ell^-$ collider at the $\sim 5\sigma$ (statistical) level. For example, when 
$A_{FB}=0$ so that only a single observable is employed to distinguish them, the cross sections in the two cases are $\simeq 10.7$ (isosinglet) and 12.1 fb (isodoublet) which, assuming an 
integrated luminosity of 5 ab$^{-1}$, corresponds to 53.5k and 60.5k events. However, a $5\sigma$ statistical difference between these two samples corresponds to only $\simeq 1160$ events. As 
$y$ increases, the sample sizes in both cases will increase and an additional observable, $A_{FB}$, also becomes of use to distingish the two scenarios. Of course, systematics may play an 
important and even dominant role here but to understand those will require a more detailed simulation study.  

This same Figure also show for comparison are the two analogs of our 
benchmarks if $E$ were just an ordinary fourth generation heavy lepton, a case where we'd expect the value of $A_{FB}$ to be non-zero. We observe that both the isosinglet and isodoublet PM 
predictions as functions of $y$ {\it always} lie very far from these 'more conventional' expectations and so will be quite easily distinguishable based upon these results.  Detailed simulation studies 
should, of course, be performed to verify these tentative conclusions.

\begin{figure}[htbp]
\centerline{\includegraphics[width=5.0in,angle=0]{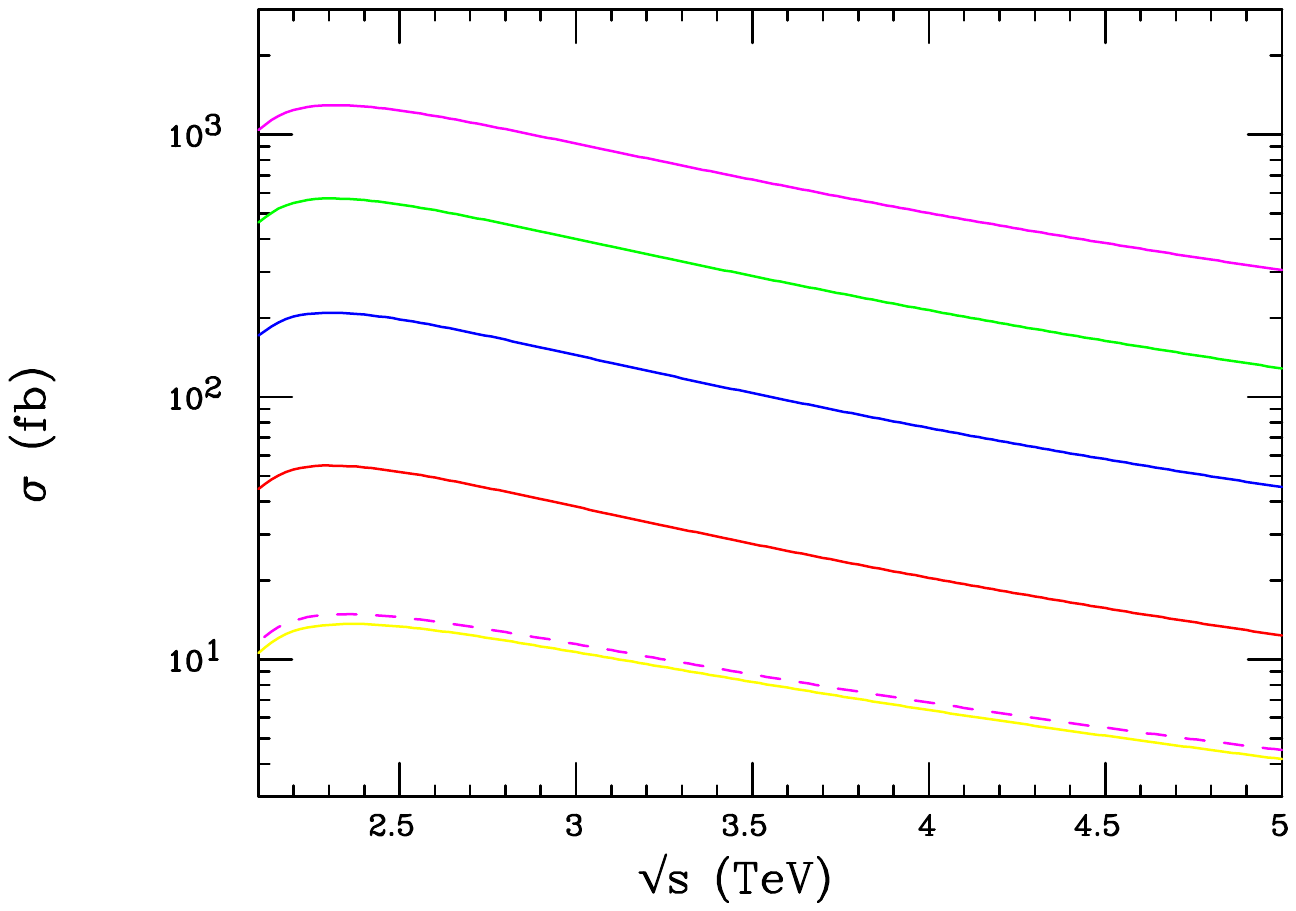}}
\vspace*{-0.8cm}
\centerline{\includegraphics[width=5.0in,angle=0]{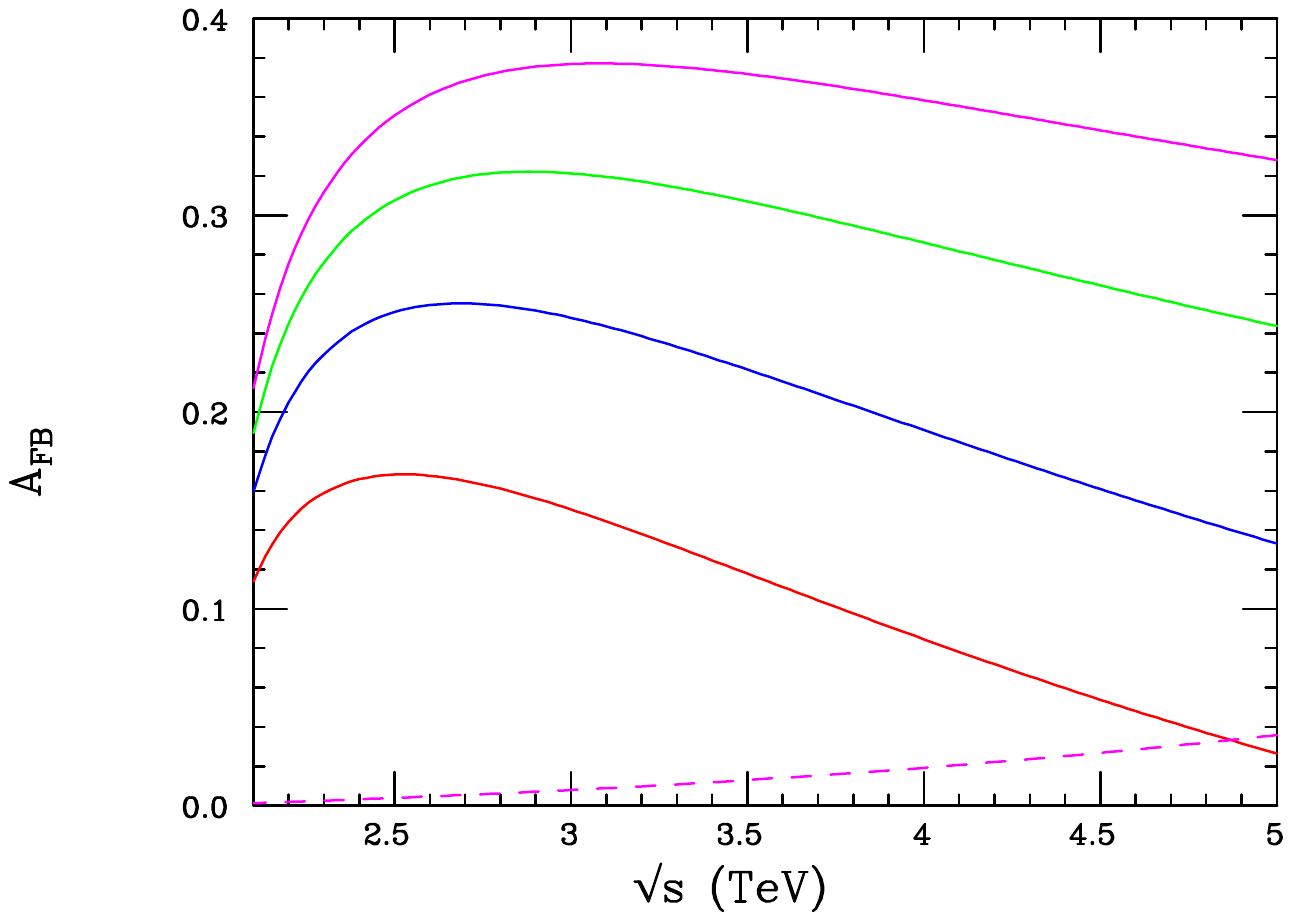}}
\vspace*{-1.3cm}
\caption{The energy dependencies of both $\sigma$ (top panel) and $A_{FB}$ (bottom panel) for isosinglet VL PM pair production in $\ell^+ \ell^-$ annihilation assuming $M_E=1$ TeV. From top 
to bottom, the curves correspond to the choices of $y=1,3/4,1/2,1/4$ and 0, in magenta, green, blue, red and yellow, respectively. Note that in the bottom panel, the prediction for the case of $y=0$ 
(in yellow), $A_{FB}=0$, lies along the horizontal axis. The dashed magenta curves in both plots correspond to the case of a 15 TeV $Z'=Z_I$ new heavy gauge boson as discussed in the text}
\label{fig4}
\end{figure}

One might ask if the dark Higgs contribution in the PM case can lead to any other visible effects on the predictions for $\sigma$ and/or $A_{FB}$ that can be used to further identify the PM lepton 
production case. We might imagine, \eg, given the additional terms in the cross section and its rather soft $\beta$ dependence arising from the purely vectorial couplings appearing in the $s$-channel 
terms, that the energy dependencies of both $\sigma$ and $A_{FB}$ may show some sensitivity to non-zero values of $y$. These possibilities are explored in both panels of Fig.~\ref{fig4}; here we see 
that this is, unfortunately, not the case. In the top panel of Fig.~\ref{fig4} we show $\sigma(\sqrt s)$ for the lighter benchmark model for different values of $y$ where we observe that the excitation 
curves are essentially parallel as $y$ is varied with relatively minor differences. The lower panel in the same Figure shows the corresponding results for $A_{FB}(\sqrt s)$ where we see a somewhat 
greater $y$ sensitivity; for larger values of $y$, the curves become somewhat flatter and the peak value of $A_{FB}(\sqrt s)$ occurs at a larger value of the center of mass energy. This case is also 
clearly different from that where there is a new heavy $Z'$, beyond the reach of the lepton collider ($\sim 10-20$ TeV, say), that is exchanged in the $s$-channel; there as $\sqrt s$ increases the 
non-SM `distortions' would naturally increase as $\sim s/M_{Z'}^2$ when the $Z'$ resonance region is approached from below. As an example of this possibility we consider the PM motivated 
weakly coupled $Z'=Z_I$ model with $m_{Z'}=15$ TeV that was introduced in Ref.~\cite{Rueter:2019wdf}; taking the parameters $g_I=g, s_I^2=0$ for purposes of demonstration we obtain the 
results now seen as the dashed curve in both panels. This is not the behavior observed here when $y\neq 0$..

Clearly, these same sets of $\sigma,A_{FB}$ data will also very trivially differentiate the present PM case from that of charged slepton pair production\cite{Freitas:2003yp} which has a somewhat 
smaller cross section and with a corresponding angular distribution, $\sim 1-z^2$, quite different from the case of VL fermion production with or without the additional $t$-channel dark Higgs 
contribution.

\subsection{Left-Right Asymmetry}

As is well-known, having polarized beams at lepton colliders provides additional opportunities to probe new physics scenarios as well as the details of the SM.  In particular, if {\it longitudinal} beam 
polarization is available at our future TeV-scale lepton collider, the left-right polarization asymmetry, $A_{LR}$, can serve as another observable to help distinguish leptonic PM production from that 
of other color singlet possibilities.  This asymmetry can be defined in a familiar, $z$-dependent manner as 
\begin{equation}
A_{LR}(z)=P_{eff}~\frac{d\sigma_L(z)-d\sigma_R(z)}{d\sigma_{L+R}(z)}\,,
\end{equation}
where the difference, $\Delta_{LR}$, 
\begin{equation}
\Delta_{LR} = {\rm d} \sigma_L - {\rm d} \sigma_R \,.
\end{equation}
in the left- and right-handed polarization-dependent cross sections in the numerator is given by 
\begin{equation}
\Delta_{LR}=\frac{\pi\alpha^2}{2s}\beta~\Bigg(P_{ij}A_{ij} [2-\beta^2(1-z^2)] -2\lambda R_iC_i ~\Big[\frac{(1-\beta z)^2+1-\beta^2}{a-\beta z}\Big]+ \Big[\frac{\lambda}{2}~ \frac{(1-\beta z)}{(a-\beta z)}\Big]^2 \Bigg)\,,
\end{equation}
and where we've now employed the coupling combination
\begin{equation}
A_{ij}=(v^e_ia^e_j+a^e_iv^e_j)v^E_iv^E_j\,,
\end{equation}
with $P_{eff}$ being the effective longitudinal beam polarization needed to generate the asymmetry. In general when both, \eg, $e^\pm$, beams are polarized, then one has
\begin{equation}
P_{eff}=\frac{P_{e^-}-P_{e^+}}{1-P_{e^-}P_{e^+}}=\frac{1.1}{1.24}\simeq 0.89\,,
\end{equation}
where here we've assumed that $P(e^-)=0.8$ and $P(e^+)=-0.3$, analogous to the canonical case of the ILC\cite{lc}, for demonstration purposes; results for other polarization values can be easily 
obtained via an overall rescaling. The corresponding angular integrated left-right asymmetry, $A_{LR}$, is obtained as usual by integrating the numerator and denominator of this expression over 
all values of $z$. 

\begin{figure}[htbp]
\centerline{\includegraphics[width=5.0in,angle=0]{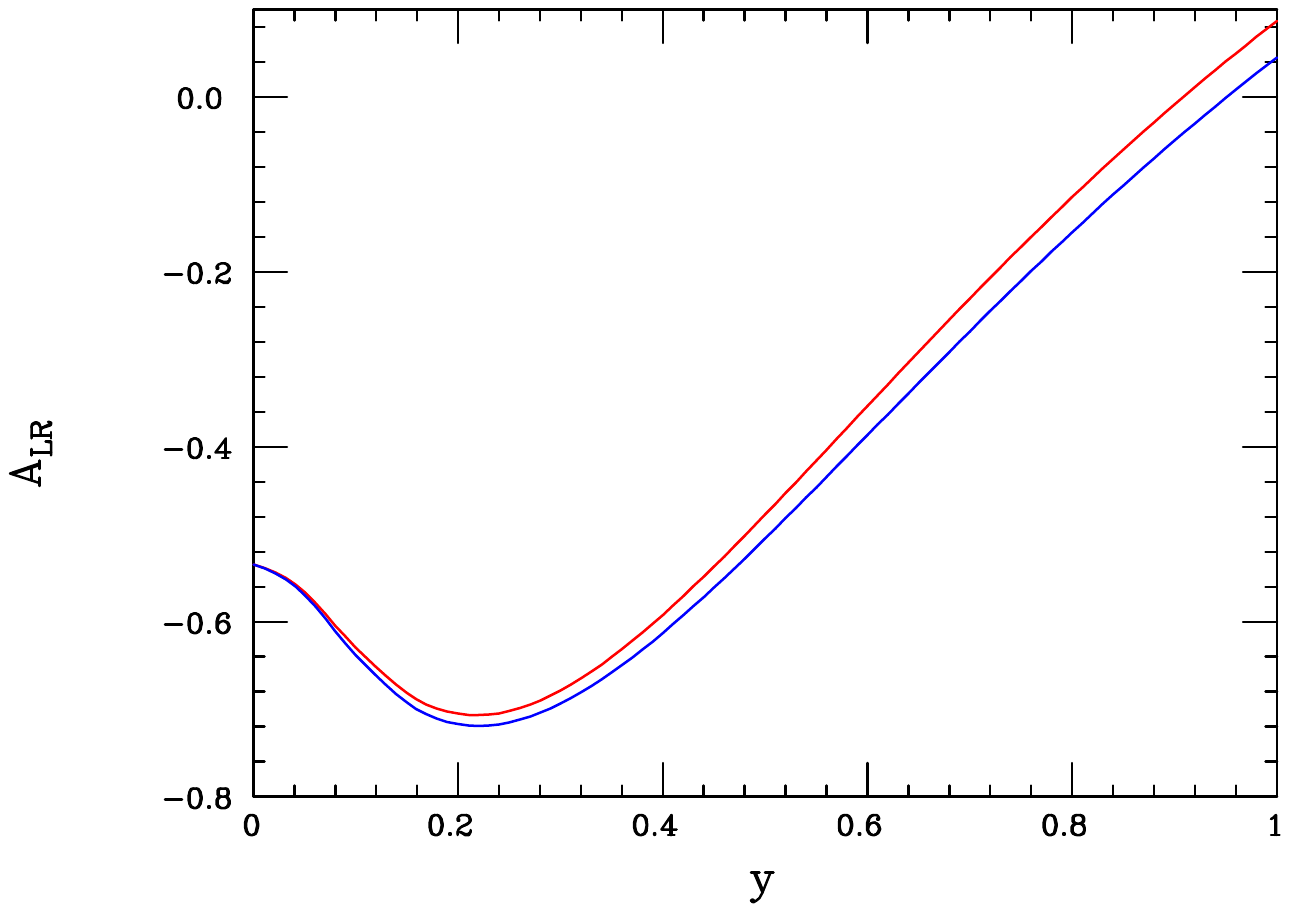}}
\vspace*{-0.8cm}
\centerline{\includegraphics[width=5.0in,angle=0]{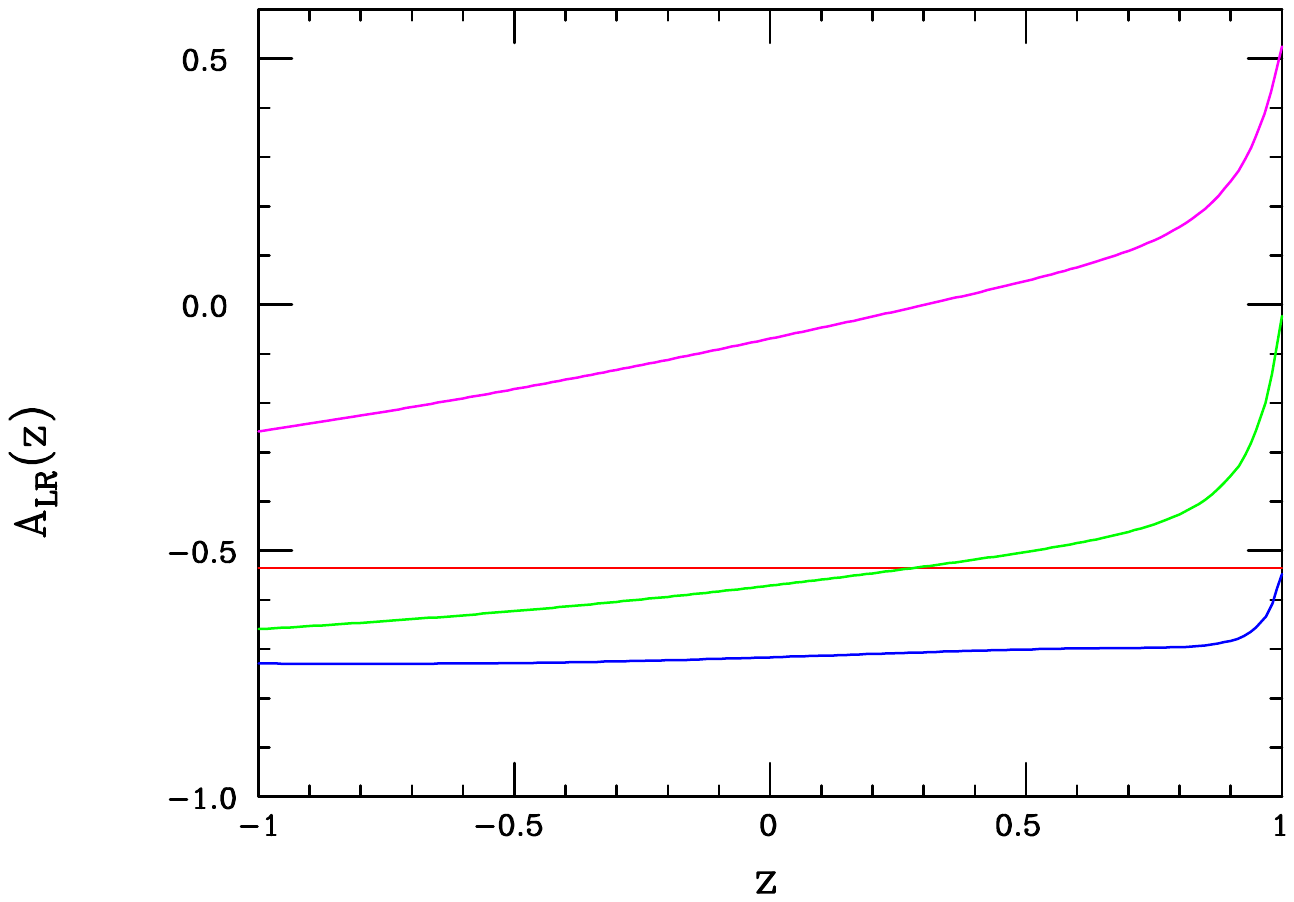}}
\vspace*{-1.3cm}
\caption{(Top) Angular averaged value of $A_{LR}$ assuming the $P_{eff}$ value described in the text for VL isosinglet PM pair production in $\ell^+ \ell^-$ annihilation as a function of $y$ 
assuming that either 
$\sqrt s=3$ TeV with $M_E=1$ TeV (BM1 top red) or $\sqrt s=5$ TeV with $M_E=2$ TeV (BM2 bottom blue) curve, respectively. (Bottom) The angular dependence of $A_{LR}(z)$, for PM production in 
$\ell^+\ell^-$ annihilation at $\sqrt s=3$ TeV with $M_E=1$ TeV (BM1) assuming that  $y=1, 0,1/2,1/4$, corresponding to the curves from top to bottom on the left-hand side of the plot,, in magenta, red, 
green and blue  respectively.}
\label{fig5}
\end{figure}

Fig.\ref{fig5} shows the numerical results for both $A_{LR}$ and $A_{LR}(z)$ employing our benchmark PM models. Here we see a rather strong sensitivity to the value of $y$, with the initial shallow 
dip in $A_{LR}$ occurring for modest $y$ values as it moves away from zero due to the destructive effect of the opposite sign of the contribution in the $s-t$-channel interference term. However, 
for much larger $y\sim 1$, we see that $A_{LR}$ 
becomes small in magnitude and is also observed to possibly change sign. When $y=0$, $A_LR(z)$ is of course flat, again due to the pure VL nature of the PM couplings to the $Z$.  However, this 
changes, albeit somewhat slowly, as a non-zero $y$ value is turned on, particularly so in the very forward,  $z\sim 1$, direction which is clearly indicative of an additional $t$-channel exchange. 
The overall upward/downward movement of the curves in each case seen here is simply the reflection of the overall $y$-dependence of the angular integrated  value of $A_{LR}$.

It is clear from these results that the behavior of $A_{LR}$ and $A_{LR}(z)$ would certainly help to distinguish our PM model scenarios from that of an ordinary VL lepton but also from the 
case, mentioned above, of a new heavy Z' under whose gauge symmetry the heavy lepton is no longer pure VL.

\subsection{Transverse Polarization Asymmetry}

If longitudinal beam polarization is not available, we may instead be able to use the transverse polarization of the beams to form the transverse polarization asymmetry, $A_T(z)$ and it's angular 
average, $A_T$, analogously to $A_{LR}(z)$ above. In the CP-conserving 
case, which interests us here, a non-zero value for this asymmetry requires both beams to be polarized simultaneously as the effective polarization is now just simply the product of that of the 
individual beam polarizations:
\begin{equation}
P_{eff}'= P_T(e^-)P_T(e^+)=(0.8)(0.3)=0.24\,,
\end{equation}
where we've again made use of the above numerical polarization values for purposes of demonstration; the results below can be easily rescaled for other values of $P_{eff}'$. To be specific, we use the 
analyses in Ref.\cite{transv} as a guide for the present setup which allows us to define the now double differential, $\phi$-dependent cross section as 
\begin{equation}
\frac{d\sigma}{dz d\phi}=T_1(z)+P_{eff}' ~T_2(z)~c_{2\phi}\,,
\end{equation}
where $c_{2\phi}=\cos 2\phi$ and $T_1(z)$ is just $(2\pi)^{-1}d\sigma/dz$, the unpolarized cross section as given above in Eq.(6) and $T_2(z)$ is given by
\begin{equation}
T_2(z)=\frac{\pi\alpha^2}{2s}\beta^3~\Bigg(P_{ij}E_{ij}(1-z^2)+2\lambda R_iD_i\Big[\frac{1-z^2}{a-\beta z}\Big]\Bigg)\,,
\end{equation}
with the following definitions for the relevant combinations of the vector and axial-vector couplings:
\begin{equation}
E_{ij}=(v^e_iv^e_j-a^e_ia^e_j)v^E_iv^E_j,~~~D_i=(v^e_i+a^e_i)v^E_i\,.
\end{equation}
It should be noted that $T_2(z)$ has an overall $\sim 1-z^2$ dependence apart from the additional $z$ dependence occurring in the denominator of the interference term which implies that it 
vanishes at $z=\pm 1$. From these equations we can then define the asymmetry, $A_T(z)$, as
\begin{equation}
A_T(z)=P_{eff}'~ \Bigg[\frac{\int_+\frac{d\sigma}{dz d\phi} - \int_- \frac{d\sigma}{dz d\phi}} {\frac{d\sigma}{dz}}\Bigg] = P_{eff}'~\Big[\frac{T_2(z)}{\pi T_1(z)}\Big]\,,
\end{equation}
where $\int_\pm$ implies integration over the regions where the values of $c_{2\phi}$ are positive or negative, respectively. It is important to note that the pure $t$-channel term does not contribute 
to the numerator of this asymmetry which will lead to an overall suppression of the asymmetery as the value of $y$ increases. As in the case of $A_{LR}$, the corresponding angular averaged 
asymmetry, $A_T$, can then 
be easily obtained, as above, be integrating the both the numerator and denominator of this expression over $z$. One thing that is immediately obvious is that since $T_2(z)$ does {\it not} have a 
piece proportional to $\lambda^2$ (unlike $T_1(z)$), we should expect that as $y$ increases $A_T$ should decrease in overall magnitude; these expectations are borne out by the results 
presented below in Fig~\ref{fig6}.

\begin{figure}[htbp]
\centerline{\includegraphics[width=5.0in,angle=0]{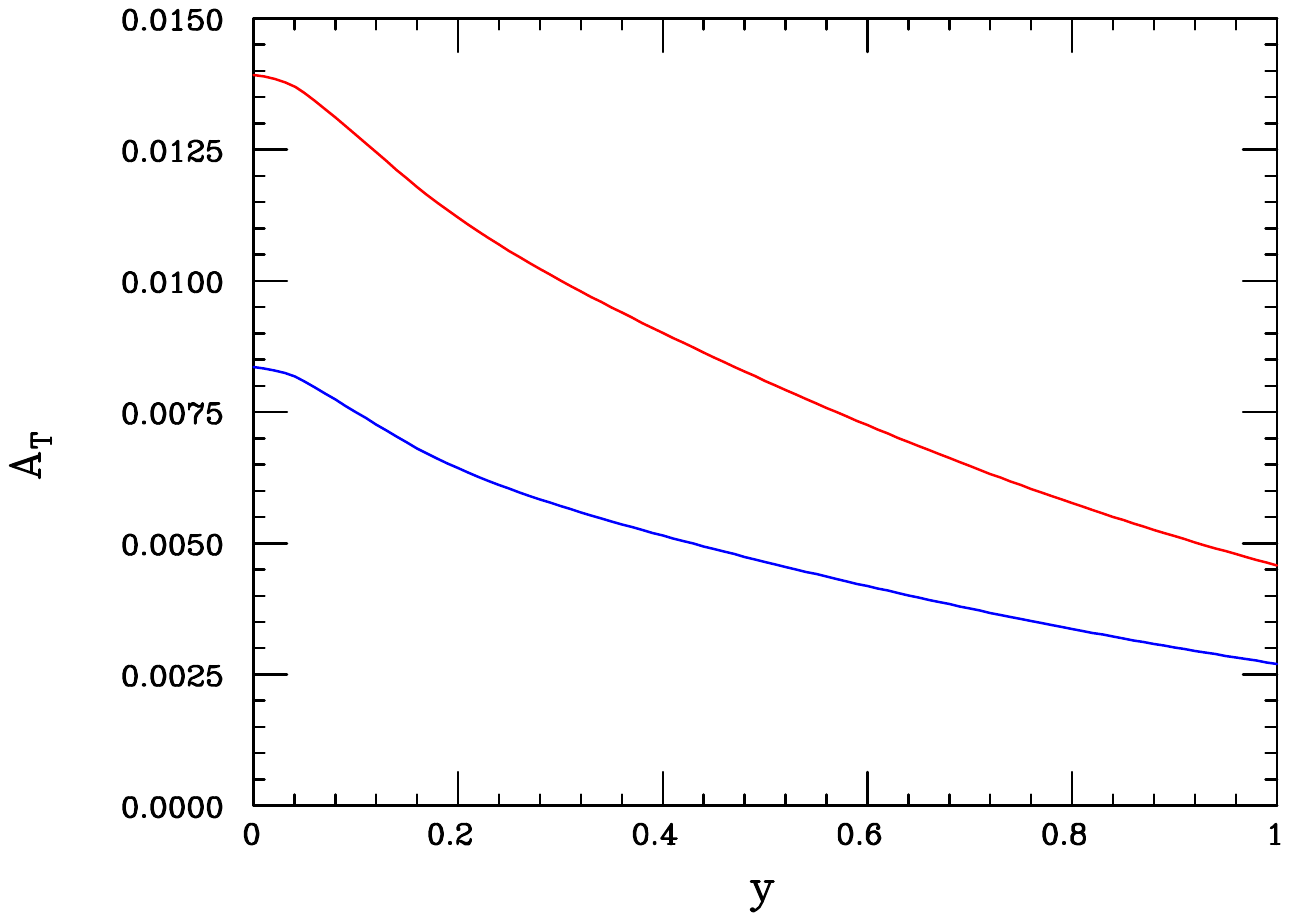}}
\vspace*{-0.8cm}
\centerline{\includegraphics[width=5.0in,angle=0]{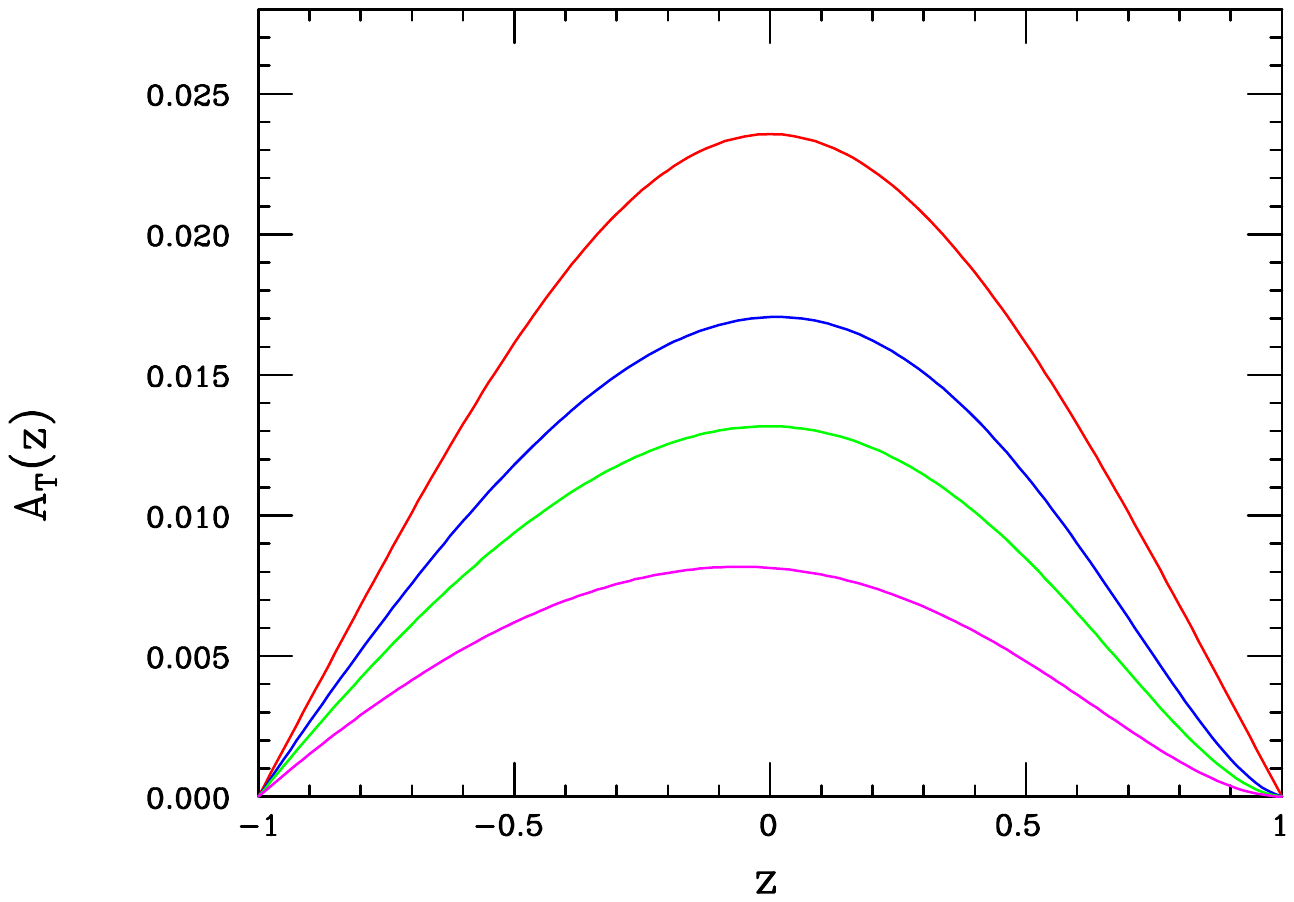}}
\vspace*{-1.3cm}
\caption{(Top) Same as the top of the panel of the previous Figure, but now for the angular averaged transverse polarization, $A_T$. (Bottom) The angular dependence of $A_T(z)$, for PM 
pair production at $\sqrt s=3$ TeV with $M_E=1$ TeV (BM1) assuming that  $y=0,1/4,1/2,1$, corresponding to the curves from top to bottom on the center of the plot in red, blue, green and magenta, respectively. In both cases we've made 
use of the $P_{eff}'$ value as described in the text.}
\label{fig6}
\end{figure}

Fig.\Ref{fig6} show both the angular dependent as well as the integrated values for $A_T(z)$ for our benchmarks scenarios. In the top panel, we see that, as expected, the overall magnitude of 
$A_T$ falls significantly as $y$ is increased and this is also reflected in the diminished height of the peak near $z\simeq 0$ in $A_T(z)$ as is seen in the lower panel in the same Figure. 
For spin-1, $s-$channel  exchanges being the sole contributor, as is the case when $y=0$, we see the anticipated SM-like $\sin^2 \theta=1-z^2$ behavior. However, for larger $y$ we observe 
a rather small asymmetry develop in this distribution towards negative values of $z$ due to the forward peaking of $t$-channel dark Higgs exchange contribution to the denominator.

\section{Like-sign PM Production in $\ell^\pm \ell^\pm$ Collisions}

\begin{figure}[htbp] 
\centerline{\includegraphics[width=5.0in,angle=0]{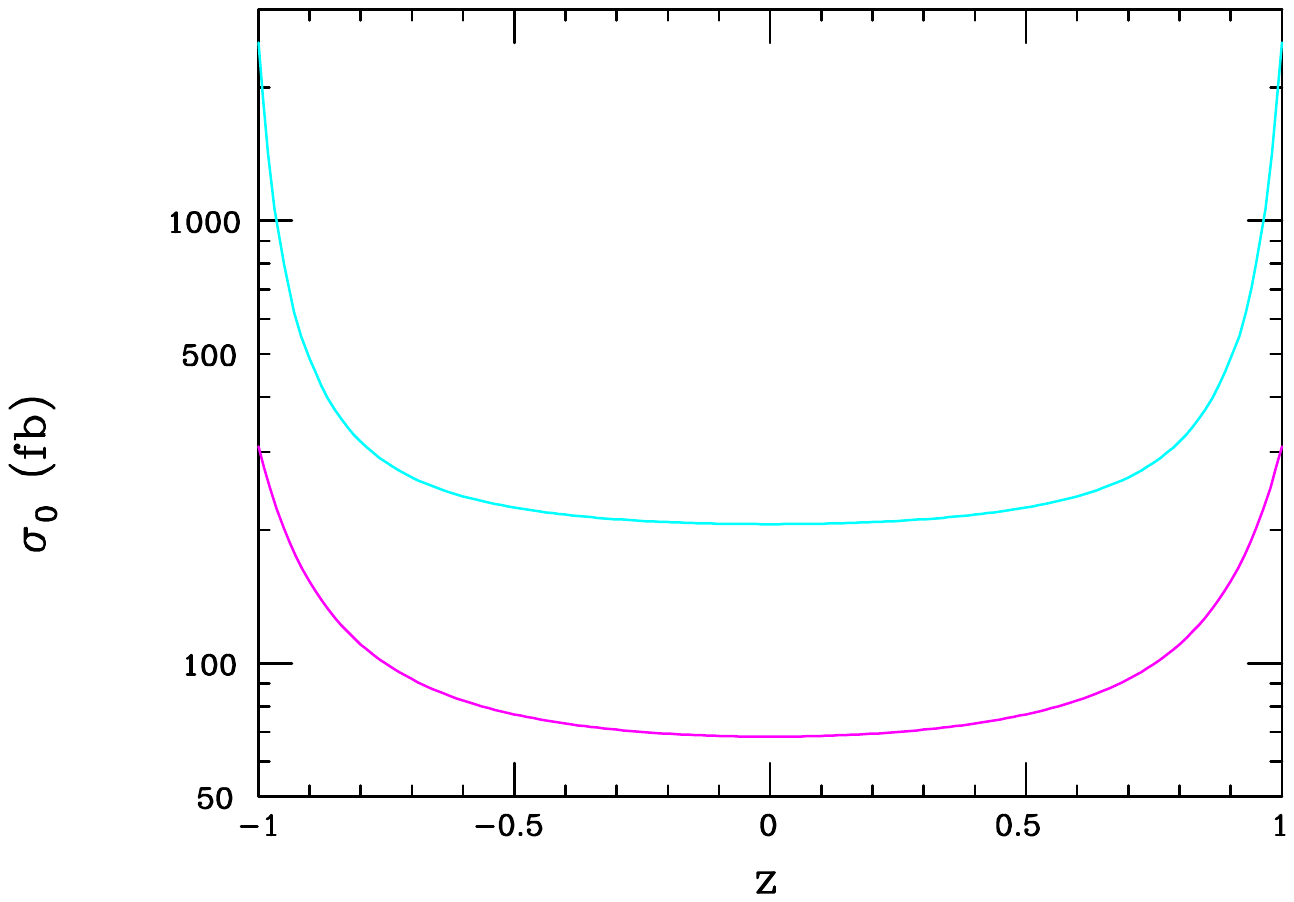}}
\vspace*{-1.3cm}
\caption{Differential like-sign PM pair production cross section, $\sigma_0(z)$, as defined in the text, for the two previously described benchmark models, $\sqrt s=3$ TeV with $M_E=1$ TeV BM1 (top 
curve) and $\sqrt s=5$ TeV with $M_E=2$ TeV (BM2 bottom curve), respectively.}
\label{fig7}
\end{figure}

In the lepton-like PM setup considered here, the obligatory decay of the PM state via its mixing with the SM analog fermion via a dark Higgs vev directly leads to the existence of other interesting 
processes which may be probed by future lepton colliders.  In particular, 
the exchange of the dark Higgs field in the $u/t$-channels will also lead to the production of like-sign VL PM fields, \ie, $\ell^\pm \ell^\pm \to E^|\pm E^\pm$, something not often considered for 
the more `ordinary' VL leptons as this might require, \eg, the existence of non-SM(-like), doubly-charged Higgs $s$-channel exchanges. This PM process is somewhat akin to Moller scattering 
except that the exchanged particle is a complex scalar and that the final state fields are quite massive. We find that the cross section for this reaction, employing the notation above and accounting for 
identical PM particles in the final state, is given by
\begin{equation}
\frac{d\sigma}{dz}=\frac{\pi\alpha^2}{16s}\beta \lambda^2~\Bigg(\Big[\frac{(1-\beta z)}{(a-\beta z)}\Big]^2+\Big[\frac{(1+\beta z)}{(a+\beta z)}\Big]^2+\frac{2}{(a^2-\beta^2z^2)}\Big[ 
(a^2-\beta^2z^2)+\frac{(1-\beta^2)^2}{4}\Big]\Bigg)\,,
\end{equation}
which we may write more simply and parametrically as 
\begin{equation}
\frac{d\sigma}{dz}=\sigma_0(z)y^4\,,
\end{equation}
with $y$ being the relevant dark Higgs Yukawa coupling. Obviously, the numerical value of this differential cross section can span {\it many} orders of magnitude due to the strong $y^4$-dependence. 
Fig.~\ref{fig7} shows the values of the function $\sigma_0(z)$ for our two benchmark models considered above; while these results appear quite large, recall that if $y=e$, \ie, $\lambda=1$, then 
the actual cross section is suppressed by a factor or $\simeq 120$ in comparison to what is shown in this Figure - still rather significant with $\sim$ ab$^{-1}$-size integrated luminosity data 
samples. Here the signature is again non-back-to-back $\ell^\pm \ell^pm$ with a large amount of missing $E_T$ and so is very clean with only rather small SM backgrounds to consider since 
lepton number emains conserved here. Note further the (obvious) property that the cross section in forward/backward symmetric as might be expected from the production of a pair of identical particles.

\section{Discussion and Conclusion}

The kinetic mixing portal for light dark matter interactions with the SM is very attractive yet necessitates the existence of new heavy fields carrying both SM and dark sector quantum numbers. The 
discovery of these new states, portal matter, will be obligatory to fully understand and verify the short distance physics associated with this portal mechanism. In particular, the certain identification of 
TeV-scale PM states at the LHC or at future colliders, especially when they are weak isospin and color-singlet VL fermions, may pose a potential problem since additional `unexpected' exchanges 
can modify the anticipated characteristics of their production process as we have seen in the discussion above. In the PM setup, these additional exchanges of a dark Higgs field are mandatory when 
the flavor of the initial state SM fermions matches that of the PM fields as these same dark Higgs are needed to allow for the PM fermions to decay. Fortunately, high energy lepton colliders are 
expected to be sufficiently clean and statistically powerful so as to provide us with enough observables to make this identification unambiguously. Even without beam polarization, the total cross 
section, $\sigma$ and the forward-backward asymmetry, $A_{FB}$, when combined in a correlated manner, were seen to be relatively powerful in differentiating the VL singlet PM scenario 
from the `vanilla' SM-like limiting case as well as from the vector-like isodoublet and 4th generation scenarios, and also those models with a new heavy $Z'$ gauge boson exchanged in the 
$s$-channel. If beam polarization is also available, then further measurements of the angular dependencies of the left-right and/or transverse polarization asymmetries, $A_{LR}/A_T$, can increase 
our confidence in these model distinctions significantly and strengthen the determination of the value for dark Higgs Yukawa coupling $y$.  Of course it is necessary that detailed simulations be 
performed to verify the strength of these conclusions.  Further, these same $t$-channel dark Higgs exchanges, present in the PM setup, will also allow for the production of the like-sign PM lepton 
final state but with a rate which is very sensitive to the magnitude of the relevant Yukawa coupling, since the cross section scales as  $\sim y^4$, between the PM and its SM analog lepton. 
However, this rather clean process is likely to be observable even for rather small values of $y\lsim 0.10-0.15$, depending upon the integrated luminosity achieved by the future lepton collider that 
we have considered.

PM plays a fundamental role in the KM scenario; its detailed nature and further implications, which can generally only be explored by colliders, are thus critical to our understanding of this mechanism.

\section*{Acknowledgements}
The author would like to particularly thank J.L. Hewett for wide ranging discussions. This work was supported by the Department of Energy, Contract DE-AC02-76SF00515.



\end{document}